\documentclass[aps,pra,twocolumn,amsfonts,nofootinbib]{revtex4-1}
\pdfoutput=1
\usepackage[utf8]{inputenc}
\usepackage[dvips]{graphicx}
\usepackage{mleftright}
\usepackage{amsfonts}
\usepackage{amssymb}
\usepackage{amsmath} 
\usepackage{bm}
\usepackage{hyperref}
\usepackage{mathrsfs} 
\usepackage{float} 
\usepackage{slashed}
\usepackage{ulem}
 
\usepackage{mathdots} 
\usepackage{braket}  
\usepackage{color} 
\usepackage{bbold}
\usepackage{anyfontsize}
\usepackage{lmodern}

\usepackage{capt-of}

\usepackage{makecell}
 
\usepackage{mathtools}

\newcounter{cquestion}

\renewcommand{\tilde}[1]{\widetilde{#1}}

\newcommand{\<}{\langle}
\renewcommand{\>}{\rangle}

\renewcommand{\cal}{\mathcal}
\newcommand{\CO}{\mathcal{O}}

\usepackage{tikz}
\usepackage{braket}
\usetikzlibrary{decorations.pathmorphing,decorations.pathreplacing}
\usetikzlibrary{decorations.markings}
\tikzset{snake it/.style={decorate, decoration=snake}}
\usetikzlibrary{arrows.meta}
\definecolor{myRed}{RGB}{150,22,22}
\newcommand{\exch}{%
  \raisebox{-0.4ex}{%
    \tikz[scale=0.7, line width=0.4pt]{
      \draw (0,0.1)--(0, -0.1);
   \draw (0,0.1) -- (0.1,0.2);
   \draw (0,0.1) -- (-0.1,0.2);
   \draw (0,-0.1) -- (0.1,-0.2);
   \draw (0,-0.1) -- (-0.1,-0.2);
    }%
  }%
}

\newcommand{\three}{%
  \raisebox{-0.4ex}{%
    \tikz[scale=0.7, line width=0.4pt]{
      \draw (0,0.15)--(0, -0.15);
   \draw (0,0.15) -- (0.12,0.2);
   \draw (0,0.15) -- (-0.12,0.2);
      \draw (0,0.15) -- (-0.12,0.1);
   \draw (0,-0.15) -- (0.12,-0.2);
      \draw (0,-0.15) -- (0.12,-0.1);
   \draw (0,-0.15) -- (-0.12,-0.2);
    }%
  }%
}

\newcommand{\contact}{%
{%
    \tikz[scale=0.7, line width=0.4pt]{
\draw (-0.1,0.1)--(0.1, -0.1);
\draw (0.1,0.1)--(-0.1, -0.1);
\filldraw (0,0) circle (0.75pt);
    }%
  }%
}

\newenvironment{monospace}{\ttfamily}{\par}

\newcounter{cexample}[section]
\numberwithin{cexample}{section}

\newcommand{\bw}{\begin{widetext}}
\newcommand{\ew}{\end{widetext}}
\newcommand{\bea}{\begin{eqnarray}}
\newcommand{\eea}{\end{eqnarray}}
\newcommand{\be}{\begin{equation}}
\newcommand{\ee}{\end{equation}}

\numberwithin{equation}{section}
\begin{document}

\title{Towards Large-Spin Effective  Theory II: \\
$O(2)$ model in $d=4-\epsilon$}

\author{Giulia Fardelli, A. Liam Fitzpatrick, Wei Li}
\affiliation{Boston University, Boston, Massachusetts 02215, USA }

\date{\today}

\begin{abstract}
We show how to construct a holographic effective theory for the leading-twist operators in the $O(2)$ model in the $4-d=\epsilon$ expansion up to $O(\epsilon^2)$, based on the separation of short-distance and long-distance effects that arises as a function of  spin $J$.  We obtain the Hamiltonian of the theory and show that it correctly reproduces all the dimensions at $O(\epsilon^2)$ of the leading twist operators  for all values of the charge $Q$ and spin $J$. The holographic Hamiltonian is given by the bulk exchange of a charged scalar $\phi$, neutral scalar $s \sim \phi \phi^*$, and a `ghost' field $c$, as well as a single local bulk interaction $(\phi \phi^*)^2$.  We analyze various aspects of the spectrum and discuss their interpretation in light of the bulk description.

 \end{abstract}

\maketitle
\tableofcontents

\section{Introduction}
Computing the Conformal Field Theory (CFT) data -- scaling dimensions and three-point function coefficients -- of strongly coupled CFTs is a wonderfully rich and difficult problem.  The rewards of determining this data are that  it fully determines the local observables of the theory and its various deformations. Several numeric approaches, including Monte Carlo, Conformal Bootstrap, and Fuzzy Sphere regulators, can accurately determine the CFT data of a large number of operators with low scaling dimensions. For generic operators with larger scaling dimensions, these methods become less efficient and eventually become intractable, but in some regimes an effective description for the higher-dimension operators emerges which can be used instead.   In this paper, we are interested in particular in the emergence of an effective description in the limit where the spin $J$ of operators is taken large.  Moreover, we want to consider states that can  intuitively be thought of as comprising multiple particles; the precise definition is that the states have charge $Q$ under an exact $O(2)$ global symmetry of the theory.

The key assumption of the construction is that at sufficiently large spin $J$ and any charge $Q \in \mathbb{N}$, the lowest-dimension states with charge $Q$ can be modeled holographically as $Q$ particles spinning around the center of AdS$_{d+1}$. If $\Phi$ is the lowest-dimension charged operator in the CFT, with charge $Q=1$, then for $Q=2$, the existence of such states spinning around in AdS can be proven using the conformal bootstrap \cite{Fitzpatrick:2012yx,Komargodski:2012ek,Pal:2022vqc,vanRees:2024xkb}.  One can moreover show that the leading interactions between such two-particle states in AdS are universally described by  long-distance tree-level exchange of bulk fields, corresponding to the lowest-twist operators in the $\Phi \times \Phi$ and $\Phi \times \Phi^*$ Operator Product Expansion (OPE). For $Q=3$, a similar result can be shown for a subset of operators \cite{CompanionPaper1}. The main intuition for this description is that a descendant of $\Phi$ with spin $J$, on kinematic grounds, rotates around AdS at a distance $\propto \ell_{\rm AdS} \log J/\Delta$ from the center.  

\begin{figure}[h]
\begin{center}
\includegraphics[width=0.1\textwidth]{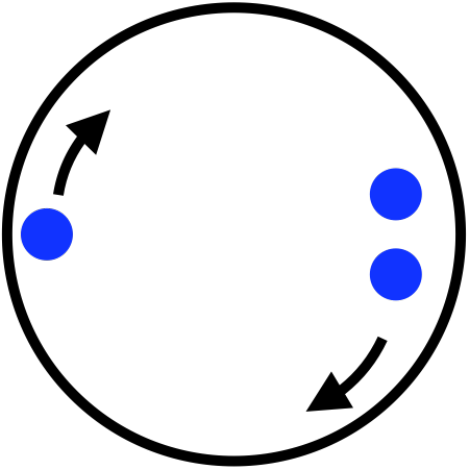}
\caption{Cartoon of the top-down view of a large spin state in AdS with three particles spinning around the center of AdS, and two of the particles close to each other near the boundary.}
\label{fig:ThreeQTwoClose}
\end{center}
\end{figure}

For $Q\ge 3$ and large $J$, one can construct a large number of $Q$-particle states in AdS with total spin $J$, such that all of the particles are far away from each other in AdS.  However, there will also be states where two or more of the particles come close to each other, as depicted in Fig.~\ref{fig:ThreeQTwoClose}, and in that case one would not expect the dimensions to be well-approximated by tree-level bulk exchanges.  Nevertheless, not all is lost, because from the perspective of the long distance $\sim \log J$, what is missing is simply an unknown short-distance interaction that becomes important on scales $\sim \ell_{\rm AdS}$.  Such short-distance interactions can easily be put in by hand in the form of local terms in an effective action, if one is provided with some (possibly nonperturbative) physical observable that can be used to fix its coefficient.  The construction then seems to be of a fairly familiar form in Effective Field Theory (EFT), where local terms are associated with the ultraviolet (UV) scale $\sim \ell_{\rm AdS}$ of the model, and long-distance effects at the infrared (IR) scale $\sim \ell_{\rm AdS} \log J$ are captured by exchanging light degrees of freedom, which generate non-local potentials. 

At a detailed level, however, it remains unclear what are the exact rules of the matching procedure when constructing the theory for a specific CFT, or if it is even possible at all.  To really test how the effective theory should work, we would like to apply it to an example where a large amount of CFT data is known at large $J$ and $Q\ge 3$.  The best candidate example for our purposes is the $O(2)$ model in the $4-d=\epsilon$  expansion.  In principle, any physical quantity in this model can be computed perturbatively to any order in $\epsilon$, and in practice the analysis of the two-loop RG equations in \cite{Kehrein:1992fn, Kehrein:1994ff,Kehrein:1995ia} provide an efficient algorithm for computing the anomalous dimensions of the leading-twist operators\footnote{The set of these operators was denoted ``${\cal C}_{\rm sym}$'' there, since they are traceless symmetric representations of  both spatial and $O(2)$ rotations.}  at any value of $Q$ and $J$ up to $O(\epsilon^2)$.  Because the central charge of this theory is not large, the bulk theory is strongly coupled already at length scales of order the AdS radius $\ell_{\rm AdS}$.  The point of the effective theory, however, is that on longer length scales the bulk theory is weakly coupled, and can be constructed  from a small subset of the CFT data. We will work out this construction for the effective Dilatation operator of the leading-twist states explicitly up to $O(\epsilon^2)$. We refer to this Dilatation operator\footnote{For different approaches on how to construct a dilatation operator at order $\epsilon$ see also~\cite{Derkachov:1995zr, DERKACHOV1995685,Liendo:2017wsn}.} as the effective Hamiltonian $H$.  At $O(\epsilon^2)$, it is a sum over one-, two-, and  three-body terms,
\begin{equation}
\begin{aligned}
H &= H_1 + H_2 + H_3, \\
H_1 & = \sum_\ell (\Delta_\Phi + \ell) a^\dagger_\ell a_\ell, \\
 H_2 &= \sum_{\ell_i} V^{(2)}_{\ell_1, \ell_2, \ell_3, \ell_4} a^\dagger_{\ell_1} a^\dagger_{\ell_2} a_{\ell_3} a_{\ell_4}, \\
  H_3 &= \sum_{\ell_i} V^{(3)}_{\ell_1, \ell_2, \ell_3, \ell_4, \ell_5, \ell_6} a^\dagger_{\ell_1} a^\dagger_{\ell_2} a^\dagger_{\ell_3} a_{\ell_4}
  a_{\ell_5} a_{\ell_6}\, ,
\end{aligned}
\end{equation} 
where $a_{\ell}$ and $a^\dagger_{\ell}$ are   annihilation and creation operators of a single-particle highest-weight state with spin $\ell$.

Remarkably, we find that with a small set of diagrams, shown in Fig.~\ref{fig:Diagrams}, the Hamiltonian of our holographic effective theory completely reproduces the anomalous dimensions of {\it all} leading-twist operators of the $O(2)$ model at $O(\epsilon^2)$ for any charge $Q$ and spin $J$. 
\tikzset{middlearrow/.style={
        decoration={markings,
            mark= at position 0.5 with {\arrow{#1}} ,
        },
        postaction={decorate}
    }
}
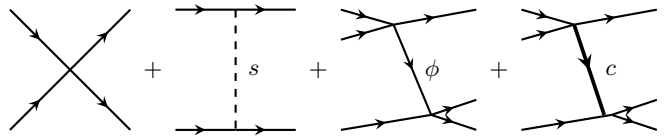
\begin{figure}[h!]
\centering
\begin{tikzpicture}
\draw[thick,middlearrow={stealth reversed}] (0.8,0.8) -- (0,0);
\draw[thick,middlearrow={stealth reversed}] (0.8,-0.8) -- (0,0);
\draw[thick,middlearrow={ stealth}] (-0.8,-0.8) -- (0,0);
\draw[thick,middlearrow={ stealth}] (-0.8,0.8) -- (0,0);
\node at (1.1,0) {$+$};
\draw[thick,middlearrow={stealth}] (1.4,0.8) -- (2.2,0.8);
\draw[thick,middlearrow={stealth}] (2.2,0.8) -- (3,0.8);
\draw[thick,middlearrow={stealth}] (1.4,-0.8) -- (2.2 ,-0.8);
\draw[thick,middlearrow={stealth}] (2.2,-0.8) -- (3,-0.8);
\draw[thick, dashed] (2.2, 0.8) -- (2.2, -0.8);
\node[right=0.05cm] at (2.2,0) {$s$};
\node at (3.3,0) {$+$};
\draw[thick,middlearrow={stealth}] (3.6,0.8) -- (4.3,0.6);
\draw[thick,middlearrow={stealth}] (3.6,0.4) -- (4.3,0.6);
\draw[thick,middlearrow={stealth}] (4.3,0.6) -- (4.8,-0.6);
\draw[thick,middlearrow={stealth}]  (4.3,0.6) -- (5.4,0.8);
\draw[thick,middlearrow={stealth}] (4.8,-0.6) -- (5.4,-0.8);
\draw[thick,middlearrow={stealth}] (4.8,-0.6) -- (5.4,-0.4);
\draw[thick,middlearrow={stealth}]  (3.6, -0.8) -- (4.8,-0.6);
\node[right=0.05cm] at (4.55, 0) {$\phi$};
\node at (5.7,0) {$+$};
\draw[thick,middlearrow={stealth}] (6,0.8) -- (6.7,0.6);
\draw[thick,middlearrow={stealth}] (6,0.4) -- (6.7,0.6);
\draw[line width=0.5mm,middlearrow={stealth}] (6.7,0.6) -- (7.1,-0.6);
\draw[thick,middlearrow={stealth}]  (6.7,0.6) -- (7.8,0.8);
\draw[thick,middlearrow={stealth}] (7.2,-0.6) -- (7.8,-0.8);
\draw[thick,middlearrow={stealth}] (7.2,-0.6) -- (7.8,-0.4);
\draw[thick,middlearrow={stealth}]  (6, -0.8) -- (7.2,-0.6);
\node[right=0.05cm] at (6.95, 0) {$c$};
\end{tikzpicture}
\caption{Diagrams for our effective Hamiltonian at $O(\epsilon^2)$. }
\label{fig:Diagrams}
\end{figure}

\section{CFT Input}
\label{sec:Input}

In order to fix the parameters of the effective theory in AdS, we need some input from the $O(2)$ model.  The first input we need is the dimension of the order parameter $\Phi$, which has \cite{Wilson:1973jj}
\begin{equation}
\Delta_\Phi = \frac{d-2}{2} + \frac{\epsilon^2}{100} +\frac{19\epsilon^3}{2000}+ O(\epsilon^4)\, .
\label{eq:DeltaPhi}
\end{equation}
We also need information about neutral operators, since they are responsible for the long-distance bulk exchanges.  The contribution of such operators is fixed by their dimensions and the OPE coefficients for them to appear in the product of $\Phi$ operators.   On general grounds, the exchange of such operators does not contribute to the Hamiltonian when their dimension is that of a Generalized Free Field (GFF) theory, i.e.~where the dimension of a composite operator is simply the sum of the dimensions of its constituents and any derivatives.  For instance, the neutral operator $S =  \Phi \Phi^*$ has dimension
\begin{equation}
\Delta_{S} = 2 \Delta_\Phi + \frac{2 \epsilon}{5} + \frac{3 \epsilon^2}{25} + O(\epsilon^3)\, ,
\label{eq:DeltaS}
\end{equation}
and consequently the leading contribution to the Hamiltonian from its exchange is at $O(\epsilon)$. At $O(\epsilon^2)$,  $\Phi$ and $S$ are the only operator exchanges that appear. To fix the coefficient of the bulk cubic coupling $s \phi \phi^*$,\footnote{Lower-case letters refer to the dual bulk field, i.e.~$\phi$ and $s$ are the bulk fields dual to the CFT operators $\Phi$ and $S$, respectively.} we also need the OPE coefficient-squared:
\begin{equation}
c^2_{S \Phi \Phi^*} = \left( 1 -\frac{2}{5} \epsilon +  O(\epsilon^2)\right)\, .
\label{eq:OpeSPhiPhi}
\end{equation}
Operators with more factors of $\Phi$ or more derivatives, including the  spin-1 current and the stress-tensor,  can be seen to contribute only at $O(\epsilon^3)$ or higher, because of suppressions coming from their OPE coefficients or dimensions, as we discuss in more detail in appendix \ref{app:HigherSpinAndTwist}.   Because the Lorentzian inversian formula~\cite{Caron-Huot:2017vep} converges down to spin $J=2$, we expect that if we include all bulk exchanges correctly then we should reproduce all the dimensions of all operators with $J\ge 2$, and so the only local terms required would be  bulk contact terms of the form $|\phi \phi^*|^n$ for integer $n$.  In a more general model, it is unlikely to be tractable to account for all such exchanges, but in the $O(2)$ model  at $O(\epsilon^2)$, the small number of operator exchanges mentioned above makes it possible to do so.  Then, all of the coefficients of the contact terms can be fixed by matching to the dimensions of the following scalar operators \cite{Kehrein:1995ia}:
\begin{equation}
\Delta_{\Phi^n} = n \Delta_\Phi + \epsilon\frac{ n(n-1)}{10} -\epsilon^2 \frac{ n (n-1)(n-3)}{50} + O(\epsilon^3).
\label{eq:DeltaPhiN}
\end{equation}

\section{Matching at $O(\epsilon)$}

At $O(\epsilon)$, the only operator exchange that has to be included is the neutral two-particle operator $S$.  In appendix \ref{app:Potentials}, we summarize the result from \cite{CompanionPaper1} for the nonlocal potential contribution to the Hamiltonian from such exchanges.  Using the dimensions of $\Phi$ and $S$ from (\ref{eq:DeltaPhi}) and (\ref{eq:DeltaS}), and the OPE coefficient from (\ref{eq:OpeSPhiPhi}), one finds that at $O(\epsilon)$ the contribution is
\begin{equation}
H_{2,S}^{\exch} = \frac{2 \epsilon}{25} \sum_{\ell_i}   \frac{\delta_{\ell_1+\ell_2, \ell_3+\ell_4}}{\ell_1+\ell_2 +1} a_{\ell_1}^\dagger a_{\ell_2}^\dagger a_{\ell_3} a_{\ell_4}  + O(\epsilon^2)\, .
\end{equation}
At this order, $H_{2,S}$ is exactly the same as the contribution from a bulk contact term. That is, a term in the  bulk  action of the form
\begin{equation}
S^{\contact} = - \int_{\rm AdS} d^{d+1} X \frac{\lambda}{4} | \phi|^4 \, , 
\end{equation}
which generates the following contribution to the leading-twist Hamiltonian:
\begin{equation}
H^{\contact} = - \frac{\lambda \epsilon^3}{80000 \pi^2} \sum_{\ell_i}   \frac{\delta_{\ell_1+\ell_2, \ell_3+\ell_4}}{\ell_1+\ell_2 +1} a_{\ell_1}^\dagger a_{\ell_2}^\dagger a_{\ell_3} a_{\ell_4}\,   .
\end{equation}
Combining both the $S$-exchange and contact-term contributions, the anomalous dimension of the charge-2 scalar operator $\Phi^2$ is
\begin{equation}
\gamma_{\Phi^2} \equiv \Delta_{\Phi^2} - 2 \Delta_\Phi = - \frac{\lambda \epsilon^3}{40000 \pi^2} + \frac{4 \epsilon}{25}\, ,
\end{equation}
which requires us to set
\begin{equation}
\lambda = -\frac{1600\pi^2}{\epsilon^2}
\end{equation}
in order to match the dimension of $\Phi^2$ from equation (\ref{eq:DeltaPhiN}). Although taking such a large and negative value of $\lambda$ naively seems disastrous, the point is that we only ever use the effective Hamiltonian on the leading-twist sector, where it is manifestly perturbative and positive; all other states have been integrated out.

The full effective Hamiltonian at $O(\epsilon)$ that we obtain from combining the exchange and contact diagrams is
\begin{equation}
\begin{aligned}
H_2 &= H^{\contact}  + H_{2,S}^{\exch}  \\
&= \frac{\epsilon}{10} \sum_{\ell_i}   \frac{\delta_{\ell_1+\ell_2, \ell_3+\ell_4}}{\ell_1+\ell_2 +1} a_{\ell_1}^\dagger a_{\ell_2}^\dagger a_{\ell_3} a_{\ell_4} +O(\epsilon^2)\, ,
\end{aligned}
\end{equation}
which exactly matches the mixing matrix $M$  in (\ref{eq:MixingMat}) at $O(\epsilon)$.  Consequently, the bulk effective Hamiltonian at $O(\epsilon)$ correctly reproduces all of the anomalous dimensions of all leading-twist operators, for any $Q$ and any $J$.

The fact that the $S$-exchange contribution at $O(\epsilon)$ is equivalent to a bulk quartic interaction simplifies many of the calculations.  We can combine both diagrams at this order into a single quartic diagram with coefficient $\tilde{\lambda}$, where
\begin{equation}
\tilde{\lambda} = 5 \lambda \, .
\end{equation}
We emphasize however that the exchange diagram at $O(\epsilon^2)$ is not equivalent to a contact term. 

As explained in the introduction, and in more detail in appendix~\ref{app:HigherSpinAndTwist},   we could have anticipated on general grounds that  at $O(\epsilon)$ the only contact term needed would be a $|\phi|^4$ term used to adjust the spin zero $Q=2$ anomalous dimension.  However, since we are only considering the leading-twist charged states ${\cal C}_{\rm sym}$, it is possible to take linear combinations of quartic bulk interactions with derivatives acting on the $\phi$s such that at $Q=2$ only the spin 0 state $\Phi^2$ is affected.\footnote{In a more general holographic setting, these interactions could be distinguished in the spectrum of $Q=2$ states by looking at subleading twists.}   In fact, in a completely general model, we expect that such additional terms would be necessary in order to adjust the OPE coefficient $c_{\Phi^2 \Phi \Phi}$.
There are two practical ways to calculate  OPE coefficients such as $c_{\Phi^2 \Phi \Phi}$ in the effective theory.  The first way relies only on using the spectrum, and the fact that these OPE coefficients show up in the large spin expansion.  For instance, at $Q=3$, the states of the form $[\Phi, \Phi^2]_J$ get an $O(\epsilon)$ anomalous dimension 
\begin{equation}
\gamma_{[\Phi, \Phi^2]_J} = \frac{\epsilon}{5} \left( 1 + \frac{2(-1)^J}{J+1} \right) + O(\epsilon^2)\, ,
\label{eq:PhiPhi2J}
\end{equation}
which at large $J$ is $\gamma_{[\Phi, \Phi^2]_J} \approx \gamma_{\Phi^2} + \frac{2\epsilon (-1)^J}{5 J}$.  The large $J$ expression~\cite{Fitzpatrick:2012yx,Simmons-Duffin:2016wlq} for the correction to $\gamma_{[\Phi, \Phi^2]_J}$ coming from $\Phi$-exchange is
\begin{align} \nonumber 
& \gamma_{[\Phi, \Phi^2]_J} - \gamma_\Phi^2 \approx -\frac{2 c_{\Phi^2 \Phi \Phi}^2 (-1)^J \Gamma \mleft(\Delta _{\Phi }\mright)^2 \Gamma \mleft(\Delta _{\Phi
   ^2}\mright)}{\Gamma \mleft(\frac{ 2 \Delta _{\Phi }-\Delta _{\Phi
   ^2}}{2}\mright) \Gamma \mleft(\frac{\Delta _{\Phi ^2}}{2}\mright)^3} \frac{1}{J^{\Delta_\Phi}} \\    \label{eq:LargeJPhiSqPhi}
   &  \approx c_{\Phi^2 \Phi \Phi}^2 \frac{ (-1)^J}{J} \left( \frac{\epsilon}{5} + \epsilon^2 \frac{\log J + \gamma_E - \frac{6}{5}}{10}\right) \, ,
\end{align}
where we have used $\Delta_\Phi$ and $\Delta_{\Phi^2}$.  Compared with (\ref{eq:PhiPhi2J}), we can read off that at leading order, $c_{\Phi^2 \Phi \Phi}^2 = 2$, as expected from the free CFT. Higher order corrections to $c_{\Phi^2 \Phi \Phi}$ can be read off in a similar way from the $Q=3$ spectrum at large $J$ at higher orders in $\epsilon$, as we will do in section \ref{sec:Q3Ep2}.

A second, more direct way to compute the OPE coefficient $c_{\Phi^2 \Phi \Phi}$ from the effective theory by computing the three-point function $\< \Phi^2 | \Phi(x) |\Phi\>$ was derived in \cite{CompanionPaper1}, with the result
\begin{align}
\nonumber
c_{\Phi^2 \Phi \Phi}^2 &= 2 +\frac{ \tilde{\lambda}  \Gamma (\Delta )^2 \left(2 H_{\Delta -1}-H_{2 \Delta
   -1}\right) \Gamma \mleft(2 \Delta -\frac{d}{2}\mright)}{4 \pi ^{d/2}\Gamma (2 \Delta ) \Gamma
   \mleft(\Delta -\frac{d-2}{2}\mright)^2} \\
    & = 2 - \frac{2\epsilon}{5} \,.
   \end{align}
\section{Matching at $O(\epsilon^2)$}

Four new effects arise when we pass to the next order in the $\epsilon$ expansion.  First, the exchange of $S$ is no longer simply proportional to a bulk contact term.  Second, we must also include exchange of the $Q=1$ operator $\Phi$.  Third, we have to adjust the coefficients of the bulk contact terms at $O(\epsilon^2)$ by matching to the CFT input data.  And fourth, at this order we must account for the fact that the bulk theory has a spurious extra primary state $\Phi^2 \Phi^*$, which in the $O(2)$ model is redundant with the descendant state $\partial^2 \Phi$ due to the equations of motion.

The first of these effects is essentially already taken into account by the expression \ref{eq:TChannelExchange} for the scalar exchange diagram; all that changes is that we have to include the higher-order powers of $\epsilon$ in the formulas (\ref{eq:DeltaS}) and (\ref{eq:OpeSPhiPhi}) for $\Delta_S$ and $c_{S \Phi \Phi^*}^2$  

The second effect requires us to evaluate the third diagram shown in Fig.~\ref{fig:Diagrams}.  This diagram is subtle because in a precise sense, it includes contributions which are second order in the contact term $H^{\contact}$, of the form $\sum_n \frac{H^{\contact} |n\rangle \langle n | H^{\contact} }{\Delta E_n}$, where the states $|n\>$ are in the conformal multiplet of the primary state created by $\Phi$.  As explained in detail in  \cite{CompanionPaper1}, one must subtract out the term where the intermediate state is the primary state itself, due to the familiar fact that in second-order time-independent perturbation theory the sum over intermediate states does not include the state being corrected.  In fact, the diagram is divergent if one does not subtract this term, since the denominator $\Delta E_n$ vanishes in that case.  After accounting for this subtlety, the diagram generates a three-body term $H_{\Phi, 3}^{\three}$.\footnote{There is also a contribution from $\Phi$-exchange where one or both of the quartic vertices come from $S$-exchange.  Fortunately, at $O(\epsilon)$, $S$-exchange is equivalent to a contact term $|\phi|^4$ interaction, so at $O(\epsilon^2)$ these contributions are completely absorbed into shifting the prefactor of $H_{\Phi, 3}^{\three}$.}  This calculation is performed in \cite{CompanionPaper1} and the result is summarized in equation (\ref{eq:ThreeToThree}). 

Accounting for these two effects, the effective Hamiltonian is a sum of the following three term:
\begin{equation}
H = H^{\contact} + H_{2,S}^{\exch} + H_{3,\Phi}^{\three}\,.
\label{eq:HAlmost}
\end{equation}
All bulk exchanges that contribute at $O(\epsilon^2)$ have been included at this point, and therefore the argument in section \ref{sec:Input} would seem to imply that all that remains is to fix the coefficients of bulk contact terms.  However, after fixing the coupling $\lambda$ to
\begin{equation}
\epsilon^3 \lambda = \pi^{d/2} \left( -1600 \epsilon + 800 (5 + \gamma_E)\epsilon^2 + O(\epsilon^3)\right)\,,
\end{equation}
where $\gamma_E \approx 0.577216$ is Euler's constant, in order to reproduce the dimension  for $\Phi^2$ at $O(\epsilon^2)$ from (\ref{eq:DeltaPhiN}), one finds that the anomalous dimensions at $Q=3$ do not exactly reproduce those from the two-loop mixing matrix (\ref{eq:MixingMat}).  For example, we can not match correctly the dimension of a family of $Q=3$ operators, we refer to as `$[\Phi, \Phi^2]_J$'.\footnote{At any integer $J$, they are constructed by taking sums over terms of the form $\partial_{\mu_1} \dots \partial_{\mu_r} \Phi \partial_{\mu_{r+1}} \dots \partial_{\mu_J} \Phi^2$ fixed in order to make a conformal primary of spin $J$.  Primaries for general $Q$ and $J$ can be constructed recursively as two-particle states formed from $\Phi$ and a $(Q-1)$ state,  following~\cite{Penedones:2010ue}.   While this procedure generates an overcomplete basis, it can be refined using the methods described in~\cite[Sec. 4.1]{Anand:2020gnn}.}
In fact,  the dimension predicted from the Hamiltonian with the three terms in (\ref{eq:HAlmost}) differs from the correct answer by
\begin{equation}
\gamma_{[\Phi, \Phi^2]_J}^{\rm (\ref{eq:HAlmost})} - \gamma_{[\Phi, \Phi^2]_J} =- \frac{\epsilon^2 (-1)^J}{(J+1)^3}\left( 1 + \frac{(-1)^J}{(J+1)} \right)\,.
\label{eq:HAlmostError}
\end{equation}
Interestingly, this error term diminishes like $\sim 1/J^3$ at large $J$, suggesting that what is missing is exchange of an operator with twist $\tau=3$.\footnote{If one  determines the same anomalous dimensions using the inversion formula~\cite{Caron-Huot:2017vep}, as in the $O(1)$ case discussed in~\cite{Bertucci:2022ptt}, there are contributions from the resummation of $[\Phi^*, \Phi^2]_{J}$ to the double discontinuity at order $\epsilon^2$, but in our framework these are  captured by the Hamiltonian terms we have already included, evaluated with lines contracted so that the composite operator  $\Phi^2$ exchanges a constituent $\Phi$. } 
In fact, though, it appears that what is going wrong is that the bulk theory is actually including an extra operator that needs to be removed.
Specifically, the operator $\Phi^2 \Phi^*$ is a twist $\tau=3$ primary operator in the free bulk theory, but as mentioned above,  by the equations of motion of the $O(2)$ model it should actually be a descendant of $\Phi$.   Another way of saying what is going wrong is that the operator $\Phi^2 \Phi^*$ appears in the bulk twice, once as a three-particle state and once as a bulk descendant of $\Phi$, and we have to remove one of these redundant extra contributions.

The presence of a bulk exchange of $\Phi^2 \Phi^*$  is not immediately obvious from the interactions we have written in (\ref{eq:HAlmost}).  The key point, though, is that in addition to the explicit bulk exchanges in our Hamiltonian, there are also automatically exchanges of composite operators.  An intuitive way to see this is by taking the contraction of $H^{\three}_{\Phi, 3}$ shown below in between the $[\Phi, \Phi^2]_J$ bra and ket state. 
\begin{align*}
\begin{tikzpicture}
\draw[thick,middlearrow={stealth}] (0,0) --(1.73205,1); 
\draw[thick,middlearrow={stealth}] (1.73205,1) -- (3.4641,2);
\draw[thick,middlearrow={stealth reversed}] (1.73205,1) -- (1.23205,1.86603);
\draw[thick,middlearrow={stealth reversed}] (2.9641, 2.86603) -- (1.23205,1.86603);
\draw[thick,middlearrow={stealth reversed}]  (1.23205,1.86603) -- (-0.500001, 0.86603);
\draw[thick,middlearrow={stealth}] (1.73205,1) -- (3.4641,0);
\draw[thick,middlearrow={stealth reversed}]  (1.23205,1.86603) -- (-0.500001, 2.86603);
\draw[myRed, dashed,line width=0.4mm] (-0.5 , 1.43301) -- (3.46 , 1.43301)  ;
\node[below=0.05cm] at (1.73205,1) {$\tilde{\lambda}$};
\node[above=0.05cm] at  (1.23205,1.86603) {$\tilde{\lambda}$};
\draw[decorate, decoration={brace, amplitude=1ex, raise=1ex}]
 (3.4641,2.86603) -- (3.4641,2)   node[pos=.5, right=2.5ex] {\footnotesize $\Phi^2$};
 \draw[decorate, decoration={brace, amplitude=1ex, raise=1ex}]
  (-0.500001, 0)  -- (-0.500001, 0.86603)  node[pos=.5, left=2.5ex] {\footnotesize $\Phi^2$};
\end{tikzpicture}
\end{align*}
Cutting through this three-to-three diagram corresponds to exchange of a non-existent $\Phi^2 \Phi^*$ state and to compensate for this spurious additional exchange, we need to remove the contribution from a dimension-three scalar exchange.  Removing such an exchange is equivalent to adding a  `ghost' field `$c$', with a bulk vertex $\sim c \phi^2 \phi^*$.  The calculation of the Hamiltonian for this term $H_{\rm gh, 3}^{\three}$ is essentially the same as that of $H_{\Phi, 3}^{\three}$, without the need to subtract the pole term with $\Delta E_n=0$.  We fix the coefficient of the coupling to the ghost by matching the dimension of the operators $[\Phi, \Phi^2]_J$.  A nontrivial test of this procedure is that a single value of the coefficient is sufficient to match all $J$. 
 The final effective Hamiltonian at $O(\epsilon^2)$ with all terms included is then
\begin{equation}
H_{\rm EFT} = H^{\contact} + H_{2,S}^{\exch} + H_{3,\Phi}^{\three} + H_{3,\rm gh}^{\three}\, .
\end{equation}

\section{Results for Spectrum}

\subsection{$Q=2$}

There is only a single leading-twist state $|[\Phi, \Phi]_J\>$ at each $J$ for $Q=2$, so in this case the anomalous dimension is simply the Hamiltonian evaluated on this state.  We find 
\begin{equation}
\begin{aligned}
\gamma_{\Phi^2} & \equiv \Delta_{\Phi^2} -2 \Delta_\Phi = \frac{\epsilon}{5} +\frac{\epsilon^2}{25}\, , \\
\gamma_{[\Phi, \Phi]_{J>0}} & \equiv \Delta_{[\Phi, \Phi]_{J>0}}  -( 2 \Delta_\Phi+J) \\ 
 & = - \frac{2 \epsilon^2}{25 J(J+1)} + O(\epsilon^3)\,,
\end{aligned}
\end{equation}
in agreement with previous results \cite{Kehrein:1995ia}.\footnote{See also~\cite{Alday:2017zzv, Henriksson:2018myn,Henriksson:2022rnm,Gopakumar:2016wkt, Gopakumar:2016cpb,Dey:2016mcs,Dey:2017oim} for modern approaches.} Note that we define the anomalous dimension of an operator ${\cal O}$ with charge $Q$ and spin $J$ as
\begin{equation}
\gamma_{\cal O} \equiv \Delta_{\cal O} - (Q \Delta_\Phi + J ) , 
\end{equation}
i.e.~as the deviation of the value of its dimension from the Generalized Free Field (GFF) theory dimension. 
\subsection{$Q=3$}
\label{sec:Q3Ep2}

\begin{figure}[h]
\begin{center}
\includegraphics[width=0.4\textwidth]{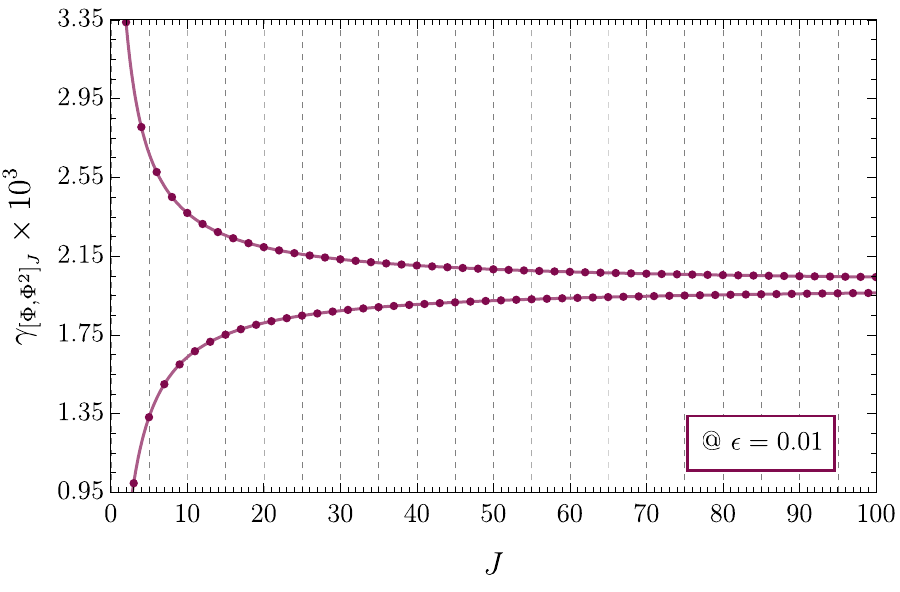}

\hspace{-0.3cm}\includegraphics[width=0.42\textwidth]{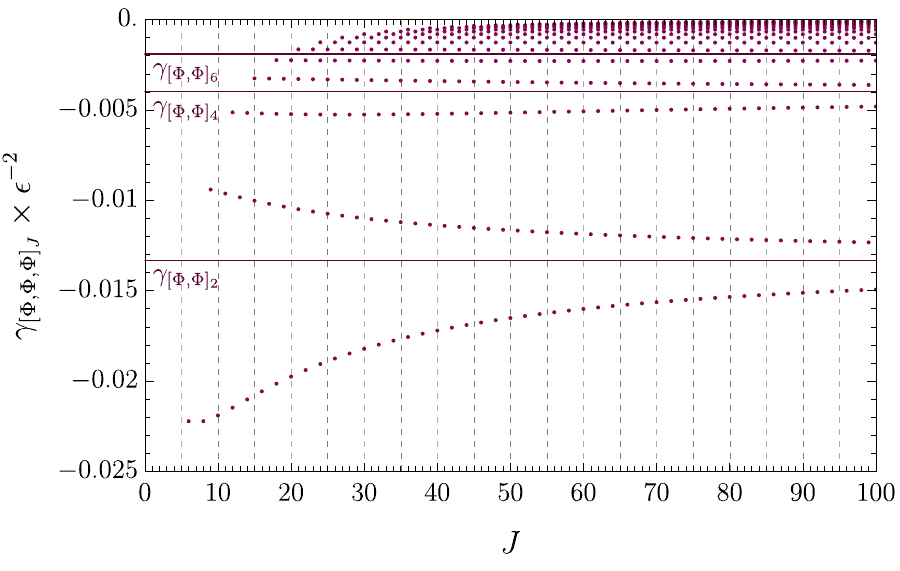}
\caption{Spectrum of $Q=3$ states, at $\epsilon = 0.01$. {\it (top)} At each spin $J$ there is a unique state, $[\Phi, \Phi^2]_J$, that receives an anomalous dimension.  {\it (bottom)} The remaining $Q=3$ eigenvalues all begin at $O(\epsilon^2)$ and are negative. }
\label{fig:Q3}
\end{center}
\end{figure}

The $Q=3$ spectrum is shown in Fig.~\ref{fig:Q3}.  At $Q=3$, only a single leading-twist state at each $J$ obtains an $O(\epsilon)$ correction to its dimension.  This state corresponds to the operator $[\Phi, \Phi^2]_J$, built from a single $\Phi$ and the scalar operator $\Phi^2$, as one might intuitively expect given that the $O(\epsilon)$ Hamiltonian in the $Q=2$ sector only affects $\Phi^2$.  Its dimension at $O(\epsilon^2)$ is
\begin{equation}
\begin{aligned}
&\gamma_{[\Phi, \Phi^2]_{J}} = \frac{\epsilon}{5} \left(1+ \frac{2 (-1)^J}{J+1}\right)  + \epsilon^2\left( \frac{1}{25} - \frac{2}{25(J+1)^3}  \right. \\
& \left. + \left\{ \begin{array}{cc} \frac{1}{5(J+1)^2} + \frac{H_J-2}{25(J+1)} + \frac{4 H_J-6}{25(J+3)} & J \textrm{ even} \\
-\frac{9}{25(J+1)^3} - \frac{H_J -4}{25(J+1)} - \frac{4H_J -4}{25(J-1)} & J \textrm{ odd} \end{array} \right\} \right),
\label{eq:GammaPhiPhiSquaredJ}
   \end{aligned}
   \end{equation}
where $H_J = \sum_{k=1}^J k^{-1}$. As a consistency check, note that in the large $J$ limit the anomalous dimension reduces to the sum $\gamma_\Phi + \gamma_{\Phi^2}$ of the anomalous dimensions of its components $\Phi$ and $\Phi^2$ (recall $\gamma_\Phi \equiv 0$ by definition).  The leading large $J$ correction is  
 \begin{equation}
 \gamma_{[\Phi, \Phi^2]_J} - \gamma_{\Phi^2} = \frac{(-1)^J}{J}\! \left(  \frac{2\epsilon}{5}  +\epsilon^2 \frac{ \log (J)+ \gamma_E -\frac{8}{5}}{5 }+ \dots \right).
 \end{equation}
Comparing this to the large spin expression (\ref{eq:LargeJPhiSqPhi}), one can read off that at $O(\epsilon)$ 
\begin{equation}
c_{\Phi^2 \Phi \Phi}^2 = 2 - \frac{2\epsilon}{5} + O(\epsilon^2)\, ,
\end{equation}
in agreement with the usual result from the $\epsilon$ expansion \cite{Dey:2016mcs}.

Although in principle the dimensions (\ref{eq:DeltaPhiN}) are needed as inputs to the theory in order to fix the contact terms, it is manifest from the structure of the loop diagrams in the CFT that at $O(\epsilon^2)$ no four-body terms are generated, so that $|\phi \phi^*|^n$ bulk terms are not present for $n\ge 4$ at $O(\epsilon^2)$. To find the coefficient of $|\phi \phi^*|^n$ for $n=3$, we match to the available CFT data. However, it turns out that at $O(\epsilon^2)$, our Hamiltonian applied to the $Q=3$ state with $J=0$ gives  $\gamma_{\Phi^3} = \frac{\epsilon}{5} + O(\epsilon^3)$ (as one can read off from (\ref{eq:GammaPhiPhiSquaredJ})),  so that we correctly match  (\ref{eq:DeltaPhiN}) at $n=3$ without the need for a bulk $|\phi|^6$ term.  We do not have a simple explanation for this apparent coincidence. 

All the remaining leading-twist states at $Q=3$ have anomalous dimensions that begin at $O(\epsilon^2)$, and are negative, as shown in Fig.~\ref{fig:Q3}.

\subsection{$Q=4$}
\begin{figure}[t]
\centering
\includegraphics[width=0.43\textwidth]{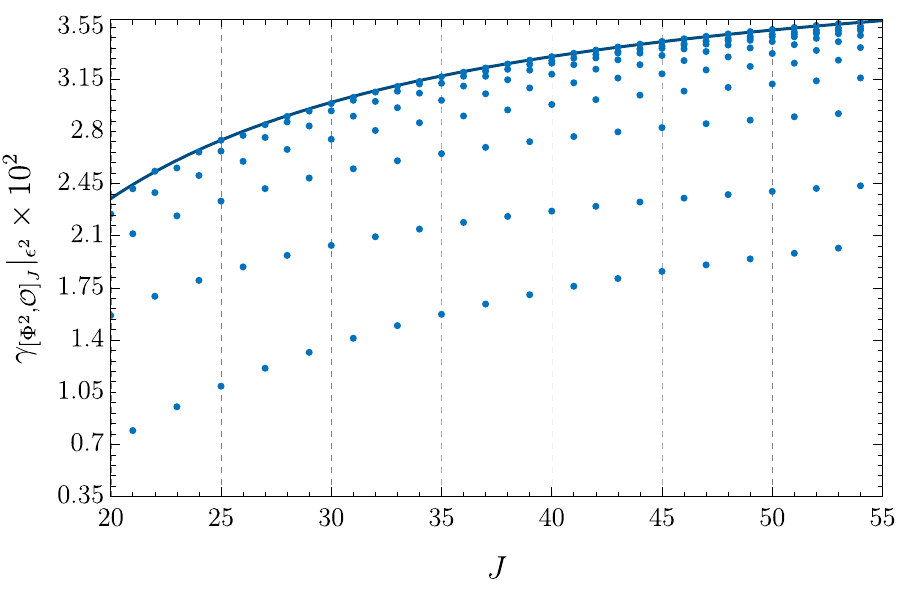}
\caption{Order $\epsilon^2$ spectrum of degenerate $Q=4$ states with $\frac{\epsilon}{5}$ anomalous dimension.  Lower trajectories are identified with $[\Phi^2, [\Phi, \Phi]_{\ell\geq 2}]_{J-\ell}$ states, while the top one (solid line) with $[\Phi^2, \Phi, \Phi]_J$. }\label{Fig:Q4eps5}
\end{figure}

At $Q=4$, there are now infinitely many states that get an anomalous dimension starting at $O(\epsilon)$, as follows from the fact that we can create a $Q=4$ operator starting with any of the $Q=3$ states of the form $[\Phi, \Phi^2]_\ell$ (with an anomalous dimension at $O(\epsilon)$) and recursively adding another $\Phi$ to get $[\Phi, [\Phi, \Phi^2]_\ell]_{J-\ell}$ or starting from $\Phi^2$.  Using the results in~\cite{DERKACHOV1995685}, one can classify all such operators with nonzero anomalous dimensions at order $\epsilon$.  Specifically, for spin $J \geq 5$, there are
\begin{align*}
\frac{J}{2}-\frac{3(1-(-1)^J)}{4}
\end{align*}
states with nonvanishing anomalous dimension at order $\epsilon$. Among these, there are $\lfloor (J - 1)/6 \rfloor$ degenerate operators for odd $J$ and $\lfloor (J - 4)/6 \rfloor$ for even $J$, all with anomalous dimension $\frac{\epsilon}{5} = \gamma_{\Phi^2}$.
As shown in Fig.~\ref{Fig:Q4eps5}, this degeneracy is fully resolved at  $O(\epsilon^2)$ and we can identify these states with operators of the form $[\Phi^2, \mathcal{O}]_J$. 

\begin{figure*}[t!]
\centering
\includegraphics[width=0.85	\textwidth]{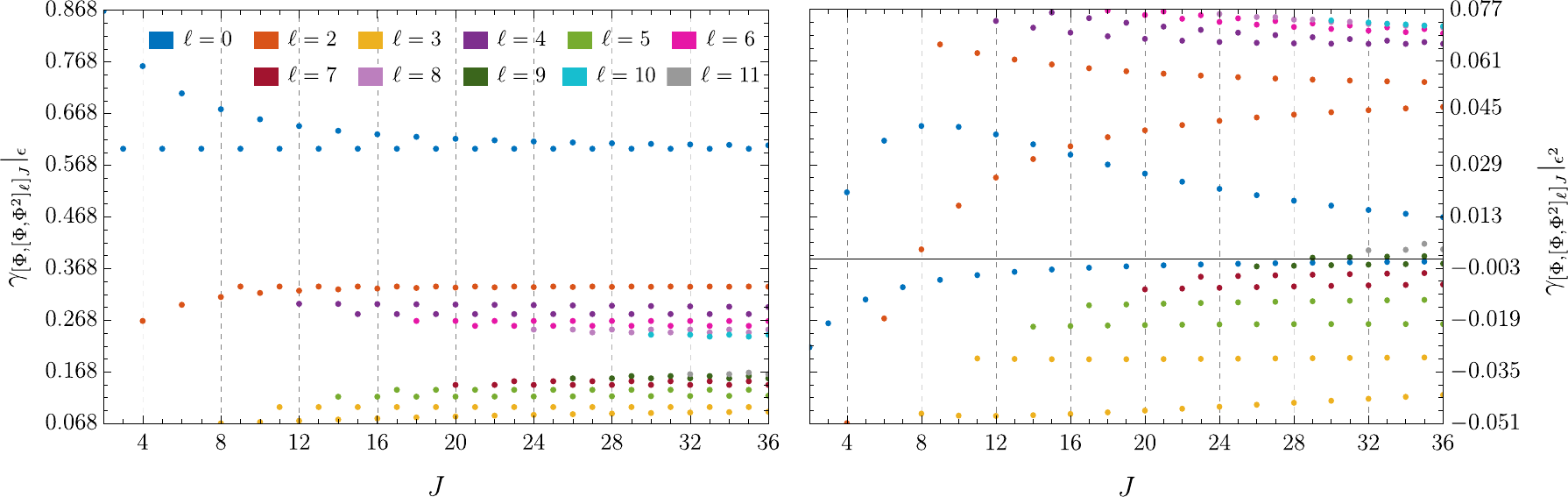}
\caption{Spectrum of $Q=4$ operators corresponding to $[\Phi, [\Phi, \Phi^2]_{\ell}]_{J-\ell}$ states at order $\epsilon$ (\textit{left}) and $\epsilon^2$ (\textit{right}). Different colors correspond to states whose anomalous dimension at order $\epsilon$ is exactly ($J$ odd) or asymptotically ($J$ even)  $\gamma_{[\Phi, \Phi^2]_{\ell}}$.}\label{fig:Q4PhiCube}
\end{figure*}

\begin{figure}[h!]
\centering
\includegraphics[width=0.43\textwidth]{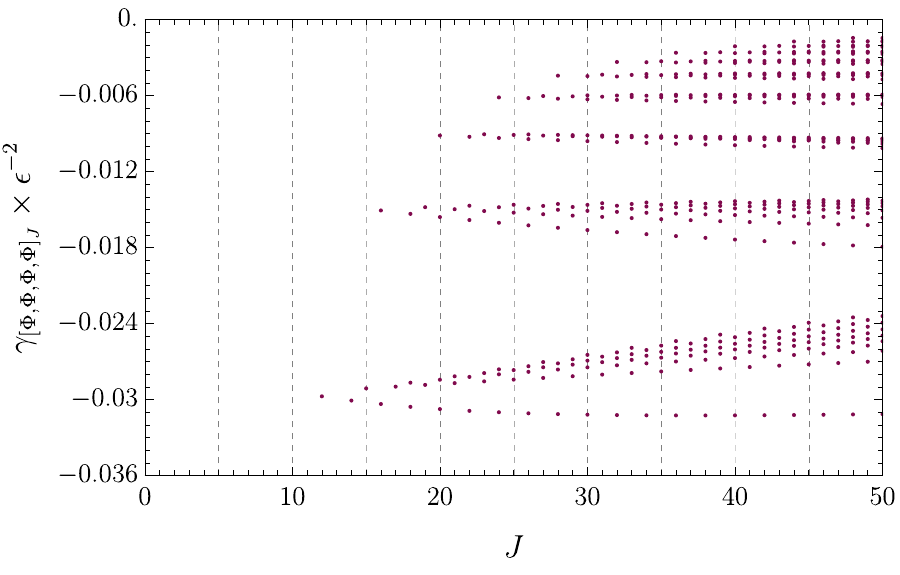}
\caption{All the $Q=4$ eigenvalues that begin at $O(\epsilon^2)$, as with $Q=3$ all of these are negative.  The lowest trajectories converge slowly to the anomalous dimensions of the $Q=2$ and $Q=3$ states starting at order $\epsilon^2$, while the four highest ones are described by the $\mathbb{Z}_4$ configuration in section~\ref{sec:ZQconfig}. }\label{fig:Q4Null}
\end{figure}

The lowest trajectories correspond to $\mathcal{O}=[\Phi, \Phi]_{\ell}$ with $\ell\geq 2 $. We interpret the highest trajectory as  $[\Phi^2, \Phi, \Phi]_J$, because at large $J$, its anomalous dimensions  asymptote to $\gamma_{\Phi^2}$. That is to say, the fact that the anomalous dimensions of the $[\Phi^2, \Phi, \Phi]_J$ operators asymptote to $\gamma_{\Phi^2}$ implies that they cannot be interpreted as a double-twist made from two lower-twist operators, but rather are in some sense `irreducibly' triple-twist.\footnote{This observation uses the fact that at $O(\epsilon^2)$, none of the $Q=2$ anomalous dimensions vanish, and none of the lowest-spin $Q=3$ anomalous dimensions are equal to $\gamma_{\Phi^2}$.} An extension of the argument in Sec.~\ref{sec:ZQconfig} should be able to  predict the subleading large‑spin behavior of this top trajectory as well,  but we leave for future investigation.
The remaining eigenvalues are not degenerate  at order $\epsilon$.  For both spin even and odd we can identify the corresponding eigenstates with $[\Phi, [\Phi, \Phi^2]_{\ell}]_{J-\ell}$,  since the anomalous dimensions are either exactly ($J$ odd) or asymptotically ($J$ even)
\begin{align}
\begin{aligned}
&\gamma_{[\Phi, \Phi^2]_{\ell}}=\frac{\epsilon}{5}\left(\! 1+\frac{2(-1)^\ell}{\ell+1}\right), \,\, \, \ell=0,2, \cdots, \ell_{\rm max}\, , \\
&\ell_{\rm max}=\begin{cases} 
2\lfloor \frac{J-1}{6}\rfloor+\text{mod}\left( \frac{J-1}{2},3\right)-1, & J\text{ odd}\,,\\
2\lfloor \frac{J}{6}\rfloor+\text{mod}\left( \frac{J(J-6)}{4},3\right), & J\text{ even}\, .
\end{cases}
\end{aligned}
\end{align}
These anomalous dimensions together with the corresponding $\epsilon^2$ term are shown in Fig.\ref{fig:Q4PhiCube}. In particular for the $\ell=0$ and $J$ odd trajectory we were able to obtain a closed form  expression\footnote{In this case it is possible to obtain a closed form expression because, for $J$ odd,  the state $\ket{[\Phi, [\Phi, \Phi^2]_0]_J}$ is a true eigenstate of the Hamiltonian. For the other $\ell$'s  and for $J$ even the eigenstates corresponds to    $\ket{[\Phi, [\Phi, \Phi^2]_\ell]_{J-\ell}}$ only at large $J$.}
\begin{equation}
\gamma_{[\Phi, \Phi^3]_J} = \frac{3\epsilon}{5} - \epsilon^2 \frac{6(2H_{J+1} - 3)}{25 (J-1)(J+4)}, \quad (J \textrm{ odd}).
\end{equation}
Finally for even spin $J$, one finds an additional eigenvalue which, in the large‑$J$ limit, approaches twice the anomalous dimension of $\Phi^2$,  $2\gamma_{\Phi^2}=\frac{2\epsilon}{5}$.  Therefore we identify this state with $[\Phi^2, \Phi^2]_J$.\footnote{For $J$ odd, this state is not allowed by Bose symmetry. } Similarly as pointed out in~\cite{Henriksson:2023cnh},  interestingly this state exhibits a $\log J/J^2$ behavior at order $\epsilon$. The exact coefficient of the $\log J/J^2$ term in $\gamma_{[\Phi^2, \Phi^2]_J}$ is difficult to compute analytically. Nonetheless  one can  see contributions of this form just from the diagonal component of the Hamiltonian, for which we give the exact analytic result in appendix \ref{sec:PhiSqPhiSq}, where we also discuss in detail how the remaining contributions at  $O(\log J/J^2)$  come from mixing with other states.  
\begin{figure}[h!]
\centering
\includegraphics[trim={3cm 0 0 0},clip, width=0.4\textwidth]{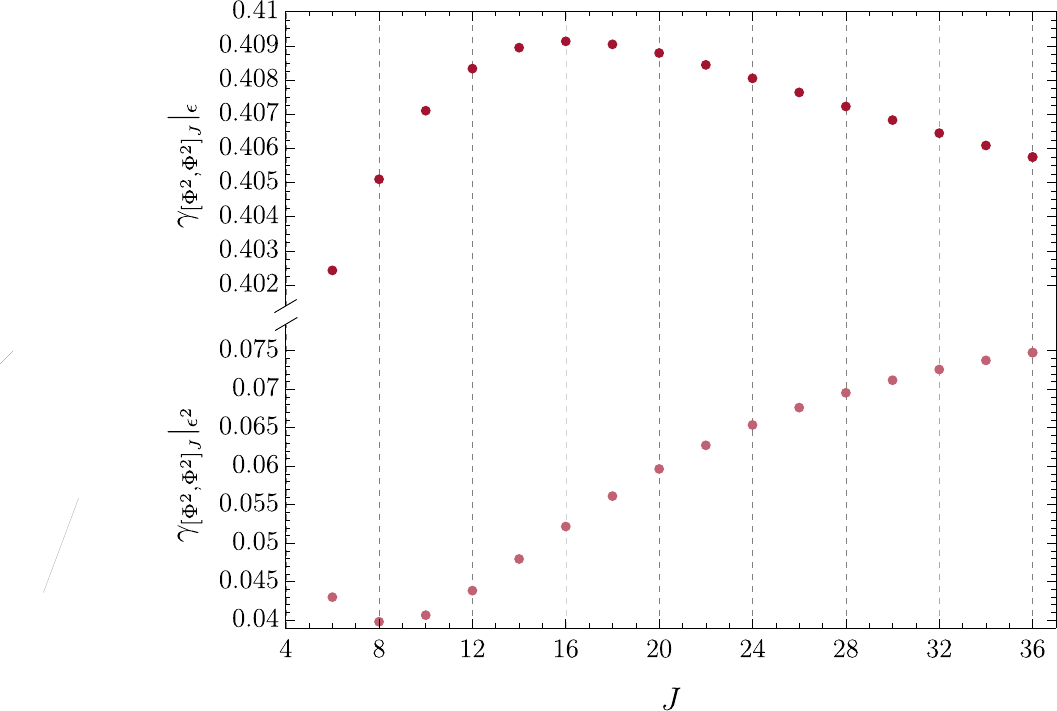}
\caption{$Q=4$ anomalous dimensions corresponding to $[\Phi^2, \Phi^2]_J$ state. The top part shows the $O(\epsilon)$ contributions and the bottom one shows the $O(\epsilon^2)$ contributions. }
\end{figure}
In Fig.~\ref{fig:Q4Null}, we show  the anomalous dimensions of all $Q=4$ operators $\CO$ where $\gamma_{\CO}$ vanishes at $O(\epsilon)$.  Compared to the corresponding figure at $Q=3$, one can now see that there are many families of accumulation points in anomalous dimension at large $J$.  

\subsection{$Q\ge 5$}
For $Q\ge 5$, although we do not have an analytic proof that our Hamiltonian completely reproduces the two-loop anomalous dimensions from \cite{Kehrein:1995ia} for all $Q$ and all $J$, we have checked a large number of $Q$ and $J$ sectors numerically, up to at least 20 digits for each eigenvalue, and in all cases found exact agreement.  
For example, in \eqref{eqn:Q10J10tb} we list all the $Q=10,J=10$ eigenvalues up at $O(\epsilon^2)$, which are the numeric values one obtains both from our effective Hamiltonian and from the two-loop calculations of \cite{Kehrein:1995ia}.
\begin{align}\label{eqn:Q10J10tb}
\begin{array}{c}
\hline\\
7.37865092050265\epsilon-8.65929359944172\epsilon^2   \\  6.91634528171365\epsilon-7.02196284307236\epsilon^2  \\   6.54312824790920\epsilon-6.26158417586321\epsilon^2  \\   6.47972100493863\epsilon-6.37713008364753\epsilon^2   \\  
6.27290895238706\epsilon-6.02175729205257\epsilon^2   \\   6.02031383266327\epsilon-5.45993463610200\epsilon^2  \\   5.75352520850278\epsilon-4.96182823034167\epsilon^2  \\
 5.60481830281109\epsilon-4.80650283519196\epsilon^2  \\  
5.28199489856803\epsilon-4.20354864652487\epsilon^2   \\
4.93187498407322\epsilon-3.60959550554365\epsilon^2  \\
4.47080206001413\epsilon-3.09891743645076\epsilon^2  \\
   \frac{8\epsilon}{3}-\frac{281210716}{324842535}\epsilon^2 \\
\\\hline
\end{array}
\end{align}

\subsection{Leading Logs}

Although in most of this paper, we limit ourselves to at most $O(\epsilon^2)$, we note that the structure of the effective Hamiltonian automatically captures certain leading large $J$ effects for any number of particles.  An example of such an effect is the leading large $J$ behavior arising from the anomalous dimension of the neutral operator $S$.
 More precisely, for any number of particles, tree-level exchange of $S$ produces a correction of the form
$c_{S \Phi \Phi^*}^2/J^{\Delta_S}$.
Using the expression from (\ref{eq:TChannelExchange}) with the all-orders anomalous dimension $\Delta_S$ automatically includes an infinite series of log-enhanced corrections proportional to
\begin{equation}
\frac{c_{S \Phi \Phi^*}^2}{J^{\Delta_S}} \supset \epsilon^2 \frac{\frac{1}{p!}(\frac{3\epsilon}{5} \log J)^p}{J^2}
\end{equation}
for all $p$, since $\Delta_S= 2-3\epsilon/5+O(\epsilon^2)$.  Moreover, these leading logarithms can only come from the correction to $\Delta_S$, and therefore are already correctly captured by the Hamiltonian matrix elements in (\ref{eq:TChannelExchange}) applied to a leading-twist state of any charge $Q$. Similar observations were made in~\cite{Gopakumar:2016wkt}.
\subsection{$\mathbb{Z}_Q$ Configuration}\label{sec:ZQconfig}
\begin{figure}[t!]
\begin{center}
\includegraphics[width=0.4\textwidth]{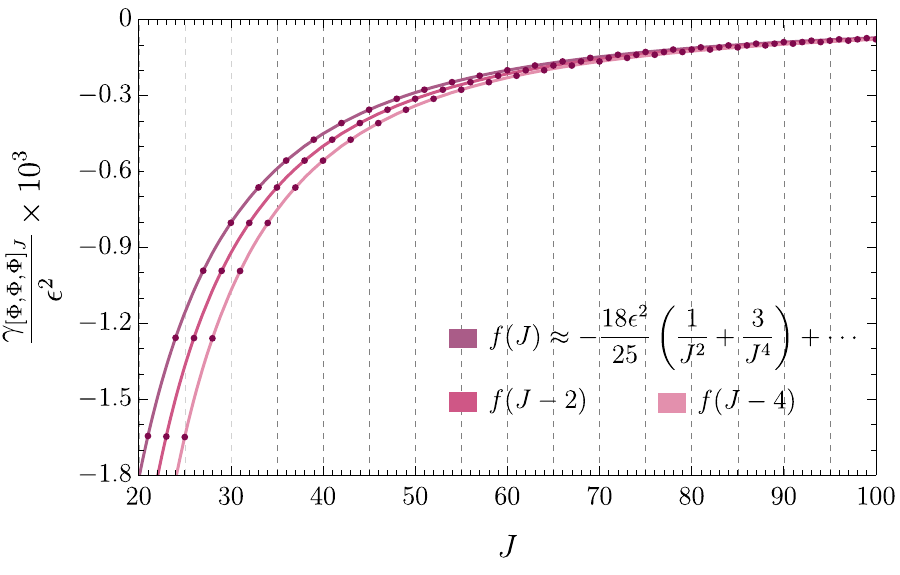}
\includegraphics[width=0.4\textwidth]{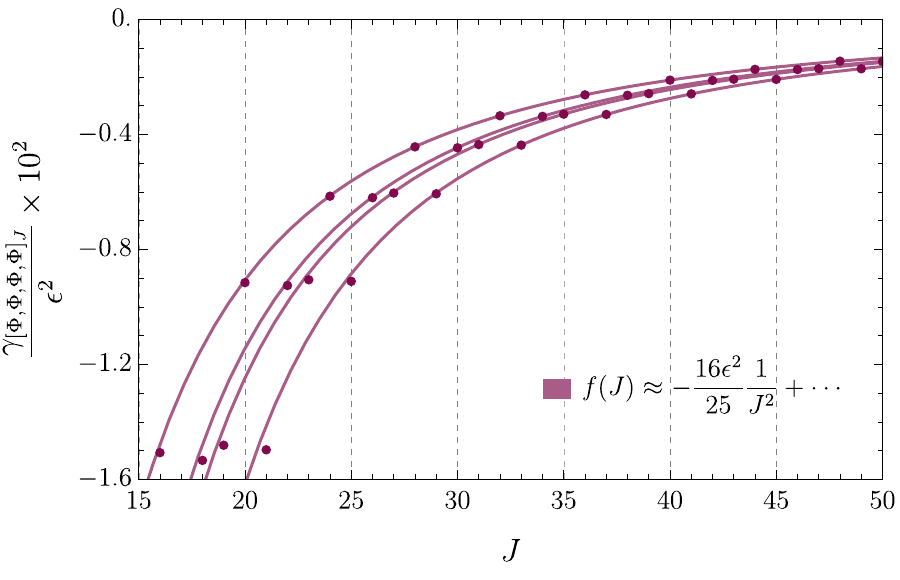}
\caption{The trajectory of eigenvalues at each spin $J$ for $Q=3$ and $Q=4$ with the smallest magnitude of their anomalous dimensions.  These states correspond to the $\mathbb{Z}_Q$ configuration described in the text, which at large $J$ are $Q$ evenly spaced particles at the boundary and have anomalous dimensions approximately given by summing over pairwise $S$ bulk exchanges.}
\label{fig:clocks}
\end{center}
\end{figure}

There is another type of large $J$ effect that is captured by tree-level $S$-exchange is the large $J$ anomalous dimension of any $Q$-particle state,  where the $Q$ particles are evenly spaced around the center of AdS, i.e.~separated in angle by $2\pi/Q$.  We will refer to this as the $\mathbb{Z}_Q$ configuration due to its discrete cyclic symmetry.  By spacing the particles evenly, this configuration keeps them as far apart as possible, and therefore the leading large $J$ anomalous dimension is just the sum over pairwise exchanges.  In Fig.~\ref{fig:clocks}, we show a blow-up of these states, which are visible in Figs.~\ref{fig:Q3} and \ref{fig:Q4Null} as the (negative) anomalous dimensions closest to zero at each $J$. As derived in \cite{Fardelli:2024heb}, the leading large $J$ anomalous dimension is
\begin{equation}
\begin{aligned}
\gamma_{\mathbb{Z}_Q}(J)& \approx \frac{Q}{2} \sum_{n=1}^{Q-1} \gamma_{[\Phi, \Phi]_{\frac{2J}{Q} \textrm{sin} \frac{\pi n}{Q}}} \\
&=  - \frac{\epsilon^2}{300}\frac{Q^3(Q^2-1)}{J^2}\, ,
\end{aligned}
\end{equation}
up to higher order corrections in $1/J$.

We can extend this prediction to $O(\epsilon^3)$ using the $Q=2$ anomalous dimensions at this order \cite{Dey:2016mcs}:
\begin{equation}
\gamma_{[\Phi, \Phi]_J} \!= \frac{-2 \epsilon^2}{25 J(J+1)} + \frac{\epsilon^3}{125}\!\left(  \frac{4 J^2 -6J-5}{J^2(J+1)^2}- \frac{6 H_J}{J(J+1)} \!\right) \, .
\end{equation}
Therefore the leading large $J$ anomalous dimension of the $\mathbb{Z}_Q$ configuration at $O(\epsilon^3)$ is
\begin{align}
\nonumber
\gamma_{\mathbb{Z}_Q} &\approx  - \frac{\epsilon^2}{300} \left(1 + \frac{3\epsilon}{5} \left( \gamma_E -\frac{2}{3} + \log\frac{2J}{Q}\right)\right) \frac{Q^3(Q^2-1)}{J^2} \\
&\quad\,  - \frac{3 Q^3 \epsilon^3 }{500 J^2} \sum_{n=1}^{Q-1} \frac{\log(\sin \frac{n \pi}{Q})}{\sin^2 \frac{n \pi}{Q}} \, .
\end{align}

\section{$\Phi^2$ as Fundamental Bulk Field}
An alternative approach one might consider is to treat $T=\Phi^2/\sqrt{2}$ as a   fundamental bulk field and study the  anomalous dimensions of double-twist operators $[\Phi, T]_J$.  In this section, we will show that simply introducing $T$ with a naive tree-level bulk interaction {\it fails} to reproduce the dimensions of $[\Phi, \Phi^2]_J$ as given in~\eqref{eq:GammaPhiPhiSquaredJ}, already at order $\epsilon$ and that fixing this discrepancy appears to require a far more involved computation than the one treating it as a composite operators that we use in the rest of this paper.  

The new setup in AdS consists of the fields  $\phi$ and $t$, with tree-level interactions of the form $\phi \phi t$, $\phi \phi s$ and $t t s$,  where the  couplings are directly determined by the corresponding OPE coefficients.  Evaluating the resulting two-body Hamiltonian on $[\Phi, T]_J$ at order $\epsilon$, as detailed in  appendix~\ref{app:HamiltonianforPhiT}, we obtain for the anomalous dimension
\begin{align}
\gamma_{[\Phi, T]_J}\equiv\Delta_{[\Phi,T]_J}-(\Delta_\Phi+\Delta_T+J)
\end{align}
due to the $\Phi$-exchange 
\begin{align}\label{eqn:Phiex}
\gamma_{[\Phi, T]_J}\Big|_\Phi =c_{\Phi\Phi T}^2(-1)^J\frac{\epsilon}{5(J+1)}\, ,
\end{align}
where $c_{\Phi\Phi T}^2\equiv c_{\Phi\Phi T^*}c_{T \Phi^*\Phi^*}$. 
The contribution from the $S$-exchange reads
\begin{align}\label{eqn:Tex}
\gamma_{[\Phi, T]_J}\Big|_S =c_{S\Phi\Phi^*}c_{STT^*}\frac{2\epsilon}{5(J+1)^2}\, ,
\end{align}
where $c_{S\Phi\Phi^*}c_{STT^*}=2-\frac{6\epsilon}{5}$. 
Comparison with our previous result in~\eqref{eq:GammaPhiPhiSquaredJ} at the same order reveals the presence of an additional contribution
\begin{equation}\label{eqn:PhiTdiff}
(\gamma_{[\Phi,\Phi^2]_J}^{\eqref{eq:GammaPhiPhiSquaredJ}}-\gamma_{\Phi^2})-\gamma_{[\Phi,T]_J}\Big|_{\Phi+S}=-\frac{4\epsilon}{5(J+1)^2}\, ,
\end{equation}
which must be canceled by including additional diagrams.  To gain intuition about the missing contributions, one can consider extracting the same anomalous dimension at  $O(\epsilon)$ using the traditional large-spin expansion~\cite{Fitzpatrick:2012yx,Komargodski:2012ek,Simmons-Duffin:2016wlq},\footnote{See also~\cite[Sec.  2.1]{CompanionPaper1}} where the relevant low-twist contributions to the OPE come from  $\Phi\times \Phi^* \rightarrow S \rightarrow T \times T^*$ and $\Phi^*\times  T \rightarrow \Phi \rightarrow \Phi \times T^*  $
\begin{align}\nonumber
\gamma_{[\Phi, T]_J} &\!\approx \frac{\epsilon}{5(J+1)}\frac{2c_{S\Phi\Phi^*}^{(0)}c_{STT^*}^{(0)}+(c_{\Phi\Phi T}^{(0)})^2(-1)^J(J+1)}{1+J+(-1)^J(c_{\Phi\Phi T}^{(0)})^2}\\
&\!\approx \epsilon\bigg[\underbrace{(c_{\Phi\Phi T}^{(0)})^2\frac{(-1)^J}{5J}\Big(1-\frac{1}{J}\Big)}_{\eqref{eqn:Phiex}}\\ \nonumber 
&\quad\, \underbrace{+\frac{2c_{S\Phi\Phi^* }^{(0)}c_{STT^*}^{(0)}}{5J^2}}_{\eqref{eqn:Tex}}\underbrace{-\frac{(c_{\Phi\Phi T}^{(0)})^4}{5J^2}}_{\eqref{eqn:PhiTdiff}}+O\Big(\frac{1}{J^3}\Big)\bigg]\, .
\end{align}
Although a direct correspondence between terms in the large-spin expansion and Hamiltonian diagrams is not entirely transparent, we can try to relate the terms in the expression above to the results in~\eqref{eqn:Phiex} and~\eqref{eqn:Tex}. The first term, whose leading behavior scales as $1/J$, can be interpreted as coming from $\Phi$-exchange, while the second term, starting at order $1/J^2$, corresponds to $S$-exchange.  Finally, the third term, which exactly cancels the second upon substituting the appropriate values of the OPE coefficients, accounts for the mismatch identified in~\eqref{eqn:PhiTdiff}.
In terms of Hamiltonian diagrams, we can interpret this contribution as arising from the one-loop diagram shown below, which is quite difficult to evaluate in AdS and we did not attempt to compute it explicitly. However, as illustrated in the figure, when $T$ is treated as a composite operator, the effect of this one-loop diagram is in fact already implicitly included from the tree-level diagrams in our original computation.
\begin{align*}
\begin{tikzpicture}
\draw[double, thick] (0,0)-- (-0.5,0.3);
\draw[double,middlearrow={stealth reversed}] (0,0)-- (-0.5,0.3);
\node[left=0.1cm, above=0cm] at (-0.5, 0.3) {\footnotesize $T$};
\node[right=0.1cm, above=0cm] at (1.5, 0.3) {\footnotesize $T$};
\draw[thick,middlearrow={stealth}] (0,0) -- (1,0);
\draw[thick, double] (1,0) -- (1.5, 0.3);
\draw[double, middlearrow={stealth}] (1,0) -- (1.5, 0.3);
\draw[thick,middlearrow={stealth}] (0,0) -- (0, -1);
\draw[thick,middlearrow={stealth}] (1,0) -- (1, -1);
\draw[thick,double] (0,-1) -- (1, -1);
\draw[double,middlearrow={stealth}] (0,-1) -- (1, -1);
\draw[thick,middlearrow={stealth}] (-0.5,-1.3) -- (0, -1);
\draw[thick,middlearrow={stealth reversed}] (1.5,-1.3) -- (1, -1);
\node[right=0.1cm, below=0cm] at (1.5, -1.3) {\footnotesize $\Phi$};
\node[left=0.1cm, below=0cm] at (-0.5, -1.3) {\footnotesize $\Phi$};
\node[myRed] at (2.6,-0.5) {$\Longleftrightarrow$};
\draw[thick,middlearrow={stealth}] (4, -1.3) --(6, -1.3);
\draw[thick,middlearrow={stealth}] (4, 0.3) --(5, 0);
\draw[thick,middlearrow={stealth}] (5, 0) --(6, 0.3);
\draw[thick,middlearrow={stealth}] (4, -1) --(5, 0);
\draw[thick,middlearrow={stealth reversed}] (5, 0) --(6, -1);
\draw[thick, myRed, dashed] (4, -0.5)-- (6, -0.5);
\draw[decorate, decoration={brace, amplitude=1ex, raise=1ex}]
  (4,-1.3) --  (4,-1)  node[pos=.5, left=2ex] {\footnotesize $T$};
  \draw[decorate, decoration={brace, amplitude=1ex, raise=1ex}]
  (6,-1) --  (6,-1.3)  node[pos=.5, right=2ex] {\footnotesize $T$};
\end{tikzpicture}
\end{align*}
\section{Application to Higher Twist}

In this paper, we have focused on the leading-twist states at charge $Q$, because at fixed $Q$ and $J$ these states are well-defined as the lowest-dimension states in their symmetry sector, with an $O(1)$ gap to higher twist states.  This separation of energies to higher twist states makes it cleaner to integrate those states out.  It would be interesting to apply these methods more generally.  The simplest possible extension is to neutral two-particle states, of the form $[\Phi, \Phi^*]_J$ which (aside from the vacuum state) are still separated from higher twist states by an $O(1)$ gap.  Consequently it is straightforward to set up the effective Hamiltonian in this sector exactly along the same lines as for the $Q=2$ states in ${\cal C}_{\rm sym}$.  Up to $O(\epsilon^2)$, the effective Hamiltonian for such states is simply given by a bulk contact term $|\phi|^4$ and scalar exchanges, now with both the neutral scalar $S$ and  the charge-2 scalar $T \equiv \Phi^2$.  The result is~\cite{Kehrein:1995ia}
\begin{equation}
\begin{aligned}
\gamma_{[\Phi, \Phi^*]_J} = -\frac{3\epsilon^2 }{25J(J+1)} , \qquad ( J > 0 )\,,
\end{aligned}
\end{equation}
and is essentially the same computation as the charged $Q=2$ sector computation, but with different OPE coefficients and combinatorics.
We leave the study of more complicated states, such as charge-1 three particle states like $[\Phi, [\Phi, \Phi^*]_J]$, to future work.

\section{Twist-to-Charge Ratio $r(Q)$}

An interesting question for any CFT with an $O(2)$ symmetry is how to describe the lowest energy state in the limit where $Q$ is large but fixed as $J$ is taken large.\footnote{More precisely, the limit is $J^{d-2} \gg Q^{d-1} \gg 1$, so the sum over pairwise exchanges of the bulk photon $\sim \frac{Q^2}{(J/Q)^{d-2}}$ is small compared to GFF energy $Q \Delta_\phi$ of $Q$ particles.} In this limit, a naive expectation is that the state should be the  $\mathbb{Z}_Q$ configuration, with $Q$ evenly spaced particles around the boundary.  However, this is not in general the lowest energy state, and at $Q>2$ it is never the lowest energy state at $O(\epsilon^2)$ in the $O(2)$ model.  The reason is that there are lower-energy configurations where some of the particles are closer to each other, so that they can take advantage of the attractive forces between them.  Instead of remaining separated, it is therefore energetically favorable at large $Q$ and $J$ for the individual particles to keep coming together into `blobs' which individually have charge $q<Q$ and spin $\ell<J$.  The optimal choice of such blobs is determined by the minimum possible value of the twist-to-charge ratio:
\begin{equation}
\begin{aligned}
r(Q) &\equiv \min_{J} \frac{\tau(Q,J)}{Q} = \Delta_\Phi + \tilde{r}(Q), \ \ 
\tilde{r}(Q) \equiv \min_J \frac{\gamma(Q,J)}{Q}\,.
\end{aligned}
\end{equation}
In the $\epsilon$ expansion, it is sufficient to work to $O(\epsilon^2)$, because already at $O(\epsilon^2)$ the lowest energy eigenvalue at every $Q$ and $J$ is negative and nondegenerate.
Therefore, at small $Q$, we can easily calculate $r(Q)$ numerically using our $O(\epsilon^2)$ Hamiltonian.  We obtain the following values: 
\begin{equation}
\begin{aligned}
 \tilde{r}(2) &= - \frac{ \epsilon^2}{150} \approx - 0.006667 \epsilon^2\, ,  && (J_{\rm min}=2) \, , \\
 \tilde{r}(3) & = - \frac{\epsilon^2}{135} \approx - 0.0074074  \epsilon^2 \, , && (J_{\rm min} = 6,8) , \\
 \tilde{r}(4) & \approx - 0.007815 \epsilon^2\,,  && (J_{\rm min} = 38)\,.
\end{aligned}
\end{equation}
For $Q=5$ we only have an upper bound, $ \tilde{r}(5) \le -0.0077452 \epsilon^2$, from computing up to $J=42$;  note that this upper bound on $\tilde{r}(5)$ is already less than the value obtained by adding $\gamma_{Q=2}$ and $\gamma_{Q=3}$, so it is a new nontrivial bound state, but it is also not yet smaller than $\tilde{r}(4)$. 
 It is  possible that at a fixed finite order in the $\epsilon$ expansion, $\tilde{r}(Q)$ is a monotonically decreasing function and a minimum value of $\tilde{r}(Q)$ is never obtained at any finite $Q$.

\section*{Acknowledgments}
We thank Gabriel Cuomo, Johan Henriksson,  Ami Katz for discussion and Johan Henriksson,  Ami Katz, Zohar Komargodski, Petr Kravchuk, and Jeremy Mann for comments on a draft. GF,  ALF, and WL  are supported by the US Department of Energy Office of Science under Award Number DE-SC0015845, and GF was partially supported by the Simons Collaboration on the Non-perturbative Bootstrap.

\appendix

\onecolumngrid

\section{Scalar Exchange Potentials}
\label{app:Potentials}
Here we summarize the results from \cite{CompanionPaper1} for the potential terms from scalar bulk exchanges.

The two-body terms from exchange of a bulk scalar with dimension $\Delta_S$ and OPE coefficient $c^2_{S \Phi \Phi^*}$ (see Fig.~\ref{fig:Diagrams}) is
\begin{equation}
\begin{aligned}
H_{2,S}^{\exch} &= \sum_{\ell_i} \delta_{\ell_1 + \ell_2, \ell_3 + \ell_4} V_S(\ell_i) a_{\ell_1}^\dagger a_{\ell_2}^\dagger a_{\ell_3} a_{\ell_4}, \\
V_{S}(\ell_i) & = \sum_{k=\frac{\Delta_S}{2}}^\infty V_S(\ell_i; k) - \sum_{k = \Delta_\Phi}^\infty V_S(\ell_i; k)\, ,\\
V_S(\ell_i; k) & = -\frac{g^2 \pi^{d/2}  }{\prod_{i=1}^4 \tilde{N}_{\Delta_\Phi, \ell_i}} a_k  \frac{ \Gamma(\Delta_\Phi)^2}{\Gamma(k)^2}\frac{2 \ell _2! \ell _4! \Gamma \mleft(\Delta _{\Phi }+k-\frac{d}{2}\mright)}{ \Gamma
   \left(\Delta _{\Phi }-k\right) \Gamma \left(\ell _2+\Delta _{\Phi }\right) \Gamma
   \left(\ell _4+\Delta _{\Phi }\right)} \\
\times \sum_{m=0}^{\ell_2} & \frac{\Gamma (k+m)\Gamma \left(k+m-\ell _2+\ell
   _4\right)   \Gamma \left(m+\ell _1+1\right)  \Gamma \left(\ell _2+\Delta _{\Phi }-k-m\right)}{\Gamma (m+1) \Gamma
   \left(\ell _2-m+1\right) \Gamma \left(m-\ell _2+\ell _4+1\right) \Gamma
   \left(k+m+\ell _1+\Delta _{\Phi }\right)} \, ,\\
\tilde{N}_{\Delta, \ell_i} &= \sqrt{ \frac{2 \pi^d \Gamma(\ell_i+1) \Gamma\mleft(\Delta - \frac{d-2}{2}\mright)}{\Gamma(\Delta+ \ell_i)}} , \\
a_k & = \frac{\Gamma (k)^2 \Gamma \mleft(\Delta _{\Phi }-\frac{\Delta _S}{2}\mright) \Gamma \mleft(\Delta
   _{\Phi }+\frac{\Delta _S}{2}-\frac{d}{2}\mright)}{4 \Gamma \left(\Delta _{\Phi }\right){}^2 \Gamma \mleft(k-\frac{\Delta _S}{2}+1\mright) \Gamma
   \mleft(k+\frac{\Delta _S}{2}-\frac{d-2}{2}\mright)} \,  ,\\
g^2 & = c_{S\Phi \Phi^* }^2 \frac{8 \pi ^{\frac{d}{2}} \Gamma(\Delta _{\Phi })^2 \Gamma \mleft(\Delta _{\Phi
   }-\frac{d-2}{2}\mright)^2 \Gamma \left(\Delta _S\right) \Gamma \mleft(\Delta _S-\frac{d-2}{2}\mright)}{\Gamma\mleft(\frac{\Delta _S}{2}\mright)^4 \Gamma \mleft(\Delta
   _{\Phi }-\frac{\Delta _S}{2}\mright)^2 \Gamma \mleft(\Delta _{\Phi } + \frac{\Delta _S-d}{2}\mright)^2}\, .
\label{eq:TChannelExchange}
\end{aligned}
\end{equation}
Expanded out to second order in $\epsilon$, 
\begin{equation}
\begin{aligned}
V_S(\ell_i)& =  \frac{2 \epsilon}{25}\frac{1}{\ell_1 + \ell_2 + 1} -\frac{\epsilon^2}{1250}\left[ 36-\frac{50}{(\ell_1+\ell_2+1)}+25 I(\ell_4\leq \ell_2) (H_{\ell_1}-H_{\ell_2}-H_{\ell_3}+H_{\ell_4})\right] + O(\epsilon^3)\, , 
\end{aligned}
\end{equation}
where $I$(true)$=1$ and $I$(false)$=-1$. 

The three-body terms from bulk exchange of $\phi$ (see Fig.~\ref{fig:Diagrams}) and the ghost field $c$ are
\begin{equation}
\begin{aligned}
H_{3,\chi }^{\three} & \sum_{\ell_i}\delta_{\ell_1+\ell_2+\ell_3, \ell_4+\ell_5+\ell_6}a^\dagger_{\ell_1}a^\dagger_{\ell_2}a^\dagger_{\ell_3}a_{\ell_4}a_{\ell_5}a_{\ell_6} V_{3,\chi}^{\three}(\ell_i)\, , \\
V_{3,\Phi}^{\three} (\ell_i)& =  \lim_{\Delta_\chi\to\Delta_{\Phi}}\left( V_{3,\Delta_\Phi}^{\three}(\ell_i, 0)-\theta(\ell_1+\ell_2-\ell_4)P_{3}^{\three}(\ell_i)  \right)+ \left(\sum_{k=1}^\infty V_{3,\Delta_\Phi}^{\three}(\ell_i, k)- \sum_{k=\Delta_{\Phi}}^\infty V_{3,\Delta_\Phi}^{\three}(\ell_i, k)\right), \\
V_{3,\rm gh}^{\three}(\ell_i) & = \left( \sum_{k=\frac{3-\Delta_{\Phi}}{2}}^\infty V_{3,\Delta_\chi=3}^{\three}(\ell_i, k)- \sum_{k=\Delta_{\Phi}}^\infty V_{3,\Delta_\chi=3}^{\three}(\ell_i, k)\right)\, ,
\label{eq:ThreeToThree}
\end{aligned}
\end{equation}
where  $\theta(x)=\begin{cases}1 &\,\, x\geq 0\, ,\\
0 & \,\,  x < 0\, ,\end{cases}\,\,$ and 
\begin{align}
&\begin{aligned}
V_{3,\chi}^{\three}(\ell_i, k) &=-\frac{g^2 \pi^{d/2}(-1)^{\ell_{12}^+-\Delta_{\Phi}+\Delta_\chi}  }{4\prod_{i=1}^6 \tilde{N}_{\Delta_\Phi, \ell_i}}\frac{\Gamma(k+1-\Delta_\Phi) \Gamma\mleft( k+2\Delta_\Phi-\frac{d}{2} \mright)\Gamma\mleft( \frac{3\Delta_\Phi-\Delta_\chi}{2}\mright)\Gamma\mleft( \frac{3\Delta_\Phi+\Delta_\chi-d}{2}\mright)}{\Gamma(\Delta_\Phi+\ell_4)\Gamma(2\Delta_\Phi+\ell_{12}^+)\Gamma\mleft(k+1+\frac{\Delta_\Phi-\Delta_\chi}{2} \mright)\Gamma\mleft(k+1+\frac{\Delta_\Phi+\Delta_\chi-d}{2} \mright)}\\
&\, \quad  \times\sum_{m=0}^{\ell_{12}^+}(-1)^m \begin{pmatrix}
\ell_{12}^+\\ m
\end{pmatrix}\begin{pmatrix} \ell_4\\ 
\ell_{12}^+ - m
\end{pmatrix} \frac{(\ell_{12}^+-m)!(\ell_3+m)! \Gamma(k+m-\ell_{12}^++\ell_4)\Gamma(k+m+\Delta_\Phi)}{\Gamma(k+m+1-\ell_{12}^+-\Delta_\Phi)\Gamma(k+m+\ell_3+2\Delta_\Phi)}\, ,
\end{aligned}\\
&\begin{aligned}
P_{3}^{\three}(\ell_i) &=\frac{1}{(\Delta_{\Phi}-\Delta_\chi )}\frac{g^2 \pi^{d/2} }{2\prod_{i=1}^6 \tilde{N}_{\Delta_\Phi, \ell_i}} \frac{(\ell_{12}^+)! (\ell_{56}^+)!\Gamma(\ell_{12}^+-\ell_4+\Delta_\chi)\Gamma\mleft(\frac{3\Delta_{\Phi}+\Delta_\chi-d}{2}\mright)^2}{(\ell_{12}^+-\ell_4)!\Gamma\mleft( \ell_{12}^++\frac{3\Delta_\Phi+\Delta_\chi}{2}\mright)\Gamma\mleft( \ell_{56}^++\frac{3\Delta_\Phi+\Delta_\chi}{2}\mright)\Gamma\mleft(\Delta_{\chi}-\frac{d-2}{2} \mright)}\,,
\end{aligned}
\end{align}
with $\ell_{ij}^+\equiv \ell_i+\ell_j$. Expanded out at second order in $\epsilon$ with $g=\tilde{\lambda}$
\begin{align}
&V_{3,\Phi}^{\three}(\ell_i) =V_{3}^{(1)}(\ell_i)+V_{3}^{(2)}(\ell_i)\, ,  \qquad \qquad  V_{3,\text{gh}}^{\three}(\ell_i) =-V_{3}^{(2)}(\ell_i)\, , \\
&V_{3}^{(1)}(\ell_i)  =\frac{2\epsilon^2}{25} \frac{  H_{\ell_1+\ell_2+\ell_3+1}+\theta(\ell_{12}^+-\ell_4)(H_{\ell_{12}^+-\ell_4}-H_{\ell_{12}^++1}-H_{\ell_{56}^++1})+\theta( \ell_4-\ell_{12}^+)(H_{\ell_4-\ell_{12}^+-1}-H_{\ell_3}-H_{\ell_4})}{(\ell_{12}^+ +1)(\ell_{56}^+ +1)}\, , \\
&\begin{aligned}
V_{3}^{(2)}(\ell_i)  &=2\epsilon^2 \sum_{k=0}^\infty\sum_{m=0}^{\ell_{12}^+}\frac{(-1)^{\ell_{12}^+-m}(m+\ell_3)! (k+m+1)! (k+m-\ell_{12}^+-\ell_4)!}{(\ell_{12}^++1)(k+1) m!(\ell_{12}^+-m)!(m-\ell_{12}^++\ell_4)!(k+m-\ell_{12}^+)!(k+m+\ell_3+2)!}\\
&\quad\, \times \left( H_{k+m+1}-H_{k+m-\ell_{12}^+}-H_{k+m+\ell_3+2}+H_{k+m-\ell_{12}^++\ell_4}-\frac{1}{k+1}\right)\,.
\end{aligned}
\end{align}
\section{Mixing Matrix at Two-Loops}
Kehrein's mixing matrix $M$ acting on ${\cal C}_{\rm sym}$ at two loops is~\cite{Kehrein:1995ia}
\begin{equation}
M = g_* V_2^{1 \rm loop} + g_*^2 (V_2^{2 \rm loop} + V_3^{2\rm loop}) + O(\epsilon^3)\, , 
\label{eq:MixingMat}
\end{equation}
where $g_* = \frac{3 }{5} \epsilon + \frac{9}{25} \epsilon^2 + O(\epsilon^3)$, and
\begin{equation}
V_2^{1 \rm loop} = \frac{1}{6} \sum_{\ell_i} \frac{1}{\ell_1+\ell_2 +1} a_{\ell_1}^\dagger a_{\ell_2}^\dagger a_{\ell_3} a_{\ell_4} \delta_{\ell_1+\ell_2, \ell_3+\ell_4}\,.
\end{equation}
\begin{equation}
\begin{aligned}
V_2^{2 \rm loop} &= -\frac{2}{9} \sum_{\ell_i} \frac{1}{\ell_1+\ell_2 +1} a_{\ell_1}^\dagger a_{\ell_2}^\dagger a_{\ell_3} a_{\ell_4} \delta_{\ell_1+\ell_2, \ell_3+\ell_4}  \\
 & \times \left( 1 - \frac{1}{2(\ell_1+\ell_2+1)} + \frac{1}{2} \theta(\ell_3 < \ell_1) (H_{\ell_3} - H_{\ell_1}) 
  + \frac{1}{2} \theta(\ell_3 > \ell_1) (H_{\ell_4} - H_{\ell_2}) \right),
\end{aligned}
\end{equation}

\begin{equation}
\begin{aligned}
V_3^{2 \rm loop} &=-\frac{1}{9} \sum_{\ell_i} \frac{ \delta_{\ell_1+\ell_2 + \ell_3, \ell_4+\ell_5+\ell_6}}{(\ell_1+\ell_2 +1)(\ell_4+\ell_5 +1)}a_{\ell_1}^\dagger a_{\ell_2}^\dagger a_{\ell_3}^\dagger a_{\ell_4} a_{\ell_5} a_{\ell_6} \Bigg[ \frac{1}{2} (H_{\ell_6} - H_{\ell_1 + \ell_2 + \ell_3 +1})  \\
 &\quad\,  \left( H_{\ell_4+\ell_5+1} - \frac{H_{\ell_4}}{2}-\frac{H_{\ell_5}}{2}\right) \theta( \ell_4+\ell_5 \ge  \ell_3) 
 + \frac{1}{2} (H_{\ell_3} - H_{\ell_6 - \ell_1 - \ell_2 -1})\theta( \ell_4+\ell_5 <  \ell_3) \Bigg] \, , 
\end{aligned}
\end{equation}
where $\theta(\rm true)=1, \theta(\rm false)=0$, and $H_\ell = \sum_{\ell=1}^k \ell^{-1}$.

\section{More about $[\Phi^2, \Phi^2]_J$}
\label{sec:PhiSqPhiSq}
\begin{figure}[t!]
  \centering
  \begin{minipage}{0.5\textwidth}
    \centering
    \includegraphics[width=0.84\textwidth]{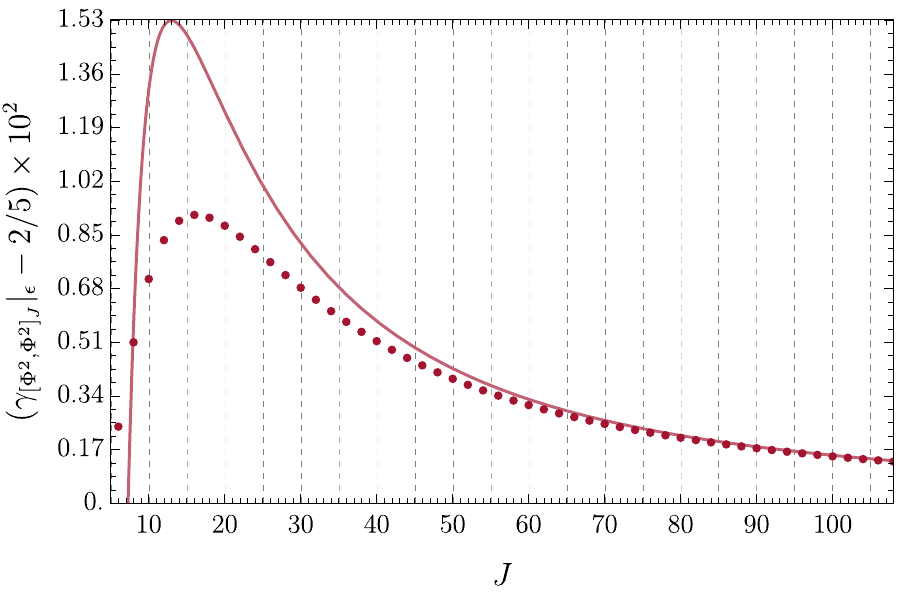}
  \end{minipage}\hfill
  \begin{minipage}{0.5\textwidth}
    \centering
    \includegraphics[width=0.86\textwidth]{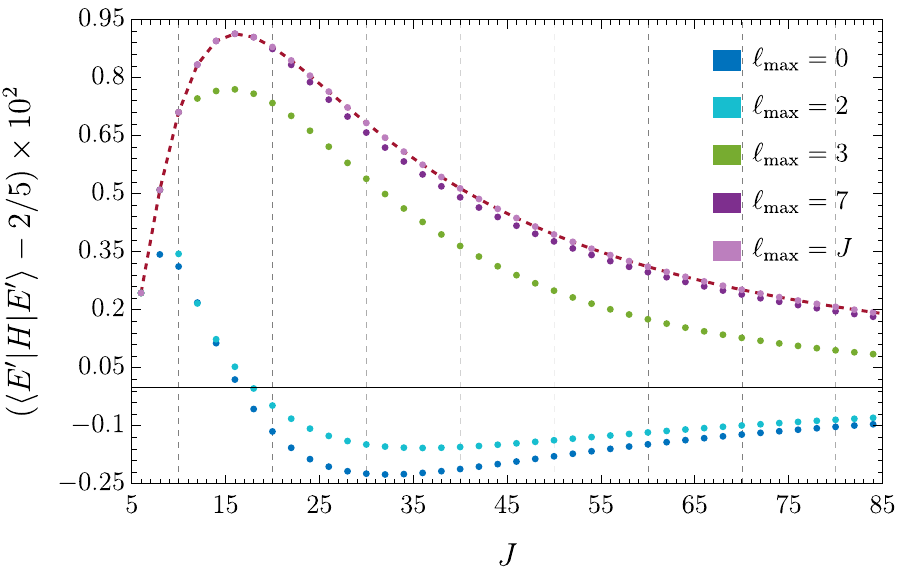}
  \end{minipage}
  \caption{Anomalous dimension at order $\epsilon$ in the $Q=4$ sector for even $J$, corresponding at large $J$ to the operator $[\Phi^2, \Phi^2]_J$. The solid line shows the fit given in~\eqref{eq:fitEpsPhi2Phi2}. \textit{(left)}. Expectation value of the order $\epsilon$ Hamiltonian for $\ket{E^\prime _{2\gamma_{\Phi^2}}}$ in~\eqref{eq:Eprime} for different truncation $\ell_{\rm max}$. The dashed line corresponds to  the exact eigenvalue,  obtained by connecting the exact data points. \textit{(right)}}
  \label{fig:Q4eps}
\end{figure}
In the main text, we observed that the $Q=4$ spectrum contains an eigenvalue which, for large even $J$, approaches twice the anomalous dimension of $\Phi^2$. This led us to identify the corresponding eigenstate $\ket{E_{2\gamma_{\Phi^2}}}$ with the double-twist operator $[\Phi^2, \Phi^2]_J$. To gain insight into the $J$-dependence of this anomalous dimension at order $\epsilon$, as a first step we can make a crude approximation by assuming $\ket{E_{2\gamma_{\Phi^2}}} \approx \ket{[\Phi^2, \Phi^2]_J}$, and study the matrix element, which we compute analytically
\begin{align}
\tilde{\lambda} \bra{[\Phi^2, \Phi^2]_J}H^{\contact} \ket{[\Phi^2, \Phi^2]_J}=\epsilon\left( \frac{2}{5}+\frac{8 \left(2 H_{J+1}+1\right)}{5 (J^2+ 3J +6)}\right)\, .
\end{align}
Notably, this expectation value already contains a ${\log J}/{J^2}$ term,
 but with a coefficient that does not quite match the one obtained by fitting the large-$J$ spectrum numerically.  See Fig.~\ref{fig:Q4eps}, where we fit the exact eigenvalue with the function
\begin{align}
\epsilon\left( \frac{2}{5}+\frac{32 H_{J+1}-88}{5 (J+1) (J+2)}\right),
\label{eq:fitEpsPhi2Phi2}
\end{align}
inspired by the structure found in~\cite{Henriksson:2023cnh}.  
The observed mismatch must come from the sum over $O(J)$ off-diagonal entries in $H^{\contact}$.  In the large-$J$ regime, these matrix elements behave as
\begin{align}
\tilde{\lambda}\langle [\Phi^2, \Phi^2]_J| H^\contact |[\Phi, [\Phi, \Phi^2]_\ell ]_{J-\ell} \rangle &\approx \frac{4\epsilon}{5} \sqrt{\frac{2(\ell+2(-1)^\ell+1)}{(J+1)(J+2)}}\left((-1)^\ell+\frac{1}{\ell+1} \right)\, .
\end{align}
Unfortunately, at large $\ell$, these elements grow like $\sqrt{\ell}$, indicating that we cannot consistently take the large-$J$ limit before summing over $\ell$. At the same time, however, performing the sum first is impractical, as obtaining analytic expressions for general $\ell$ at finite $J$ becomes increasingly difficult. For completeness, here  we  report  the other diagonal matrix elements at large $J$
\begin{align}
\tilde{\lambda}\langle [\Phi, [\Phi, \Phi^2]_\ell ]_{J-\ell}| H^\contact |[\Phi, [\Phi, \Phi^2]_\ell ]_{J-\ell} \rangle & \approx \frac{\epsilon}{5} \left( 1+\frac{2(-1)^\ell}{\ell+1}\right)+\frac{4 \epsilon (-1)^{\ell } (\ell +1) \left(\ell +5 (-1)^{\ell }+1\right)}{5 (J+1)(J+2)}\, .
\end{align}
To further support the claim that restricting  to the subspace spanned by $\ket{[\Phi^2, \Phi^2]_J}$ and $\ket{[\Phi, [\Phi, \Phi^2]_{\ell}]_{J-\ell}}$ is sufficient to capture the correct anomalous dimension, we can construct a truncated eigenstate:
\begin{align}\label{eq:Eprime}
\ket{E_{2\gamma_{\Phi^2}}^\prime} = \braket{E_{2\gamma_{\Phi^2}}|[\Phi^2, \Phi^2]_J}\ket{[\Phi^2, \Phi^2]_J}+\sum_{\ell}^{\ell_{\rm max}}\braket{E_{2\gamma_{\Phi^2}}|[\Phi, [\Phi,\Phi^2]_{\ell}]_{J-\ell}}\ket{[\Phi, [\Phi,\Phi^2]_{\ell}]_{J-\ell}}\, .
\end{align}
In Fig.~\ref{fig:Q4eps}, we plot the expectation value  of $\tilde{\lambda}\langle {E_{2\gamma_{\Phi^2}}^\prime} | H^\contact |{E_{2\gamma_{\Phi^2}}^\prime}\rangle$ at order $\epsilon$ as a function of $J$ for different values of $\ell_{\rm max}$. As expected, including a sufficient number of intermediate states, i.e.   $\ell \sim O(J/2)$, allows us to accurately reproduce the exact eigenvalue.

To conclude, we point out that every time we compute the expectation value of $H^{\contact}$ between states of the form $\ket{[\Phi^2, [\Phi, \Phi]_{\ell}]_{J-\ell}}$, we consistently find a contribution proportional to $\log J / J^2$. Importantly, the appearance of this term at order $\epsilon$ is only possible because we treat $[\Phi, \Phi]_{\ell}$ as a composite operator. At large $J$, the corresponding matrix elements take the form:
\begin{align}
\begin{aligned}
\tilde{\lambda}\langle [\Phi^2,\Phi^2]_J|H^\contact | [\Phi^2, [\Phi, \Phi]_\ell]_{J-\ell} \rangle &\approx \epsilon \frac{16\sqrt{2(2\ell+1)}}{5} \frac{H_{J+1}}{(J+1)(J+2)}\, ,\\
\tilde{\lambda }\langle [\Phi^2,[\Phi,\Phi]_\ell]_{J-\ell}| H^\contact | [\Phi^2, [\Phi, \Phi]_\ell]_{J-\ell} \rangle &\approx \frac{\epsilon}{5} + \frac{32 \epsilon(2\ell+1)}{5} \frac{H_{J+1}}{(J+1)(J+2)}\, .
\end{aligned}
\end{align}
Eventually, upon diagonalizing the full Hamiltonian, all such states flow to the same exact eigenvalue $\epsilon/5$. 
\section{Suppression of Higher Spin and Higher Twist Exchanges}
\label{app:HigherSpinAndTwist}
Using the general formula for the OPE coefficients in a GFF theory \cite{Fitzpatrick:2011dm}, and the dimension $\Delta_\Phi$ up to $O(\epsilon^2)$, it is straightforward to calculate that the OPE coefficients for double-twist operators behave at small $\epsilon$ as follows:
\begin{equation}
\begin{aligned}
c^2_{[\Phi , \Phi^*]_{n,\ell} \Phi \Phi^*} \sim \left\{ \begin{array}{cc} \epsilon^0 & n = 0 \\
   \epsilon^2 & n=1 \\
   \epsilon^4 & n\ge 2 \end{array} \right. .
\end{aligned}
\end{equation}
Moreover, the anomalous dimensions for all spinning operators require at least two insertions of the vertex $|\phi^4|$, so only the scalar double-twist operators obtain anomalous dimensions at $O(\epsilon)$:
\begin{equation}
\begin{aligned}
\gamma_{[\Phi, \Phi^*]_{n,\ell}} \sim \left\{ \begin{array}{cc} \epsilon^1 & \ell =0 \\
\epsilon^2 & \ell > 0 \end{array} \right. .
\end{aligned}
\end{equation}
These two facts, together with the large spin formula for anomalous dimensions, imply that the only double-twist exchange that contributes at $O(\epsilon^2)$ is the $n=0, \ell=0$ operator $S = \Phi \Phi^*$. That is, the contribution to the $Q=2$ states at large spin $J$ from exchange of an operator ${\cal O}$ with twist $\tau$ and spin $\ell$ is
\begin{equation}
\gamma_{[\Phi, \Phi]_J} \approx c_{\Phi \Phi^* {\cal O}}^2 \frac{2 \Gamma(\Delta_\Phi)^2 \Gamma(\tau+2\ell)}{\Gamma(\Delta_\Phi - \frac{\tau}{2})^2 \Gamma(\frac{\tau +2\ell}{2})^2 } \frac{1}{J^{\tau} }  =   c_{\Phi \Phi^* {\cal O}}^2 \times O((\tau - 2 \Delta_\Phi)^2) \,.
\end{equation}
Since the anomalous dimensions $\tau-2 \Delta_\phi$ of spinning double-twists are at least $O(\epsilon^2)$, their exchange does not contribute to the two-body terms in the Hamiltonian until $O(\epsilon^4)$.  That leaves the scalars $\ell=0$, which due to the smallness of their OPE coefficients and anomalous dimensions begin at $O(\epsilon^2), O(\epsilon^4)$ and $O(\epsilon^6)$ for $n=0, 1, \ge 2$, respectively.  Operators ${\cal O}$ with  more factors of $\phi$ are suppressed by the OPE coefficients $c_{\Phi \Phi^* \cal{O}}$, which must begin at $O(\epsilon^2)$ or higher because a single insertion of the quartic coupling in the boundary theory cannot create a connected diagram with two incoming $\Phi$ lines and $n>2$ outgoing $\Phi$ lines. 
\section{Derivation of \eqref{eqn:Phiex} and \eqref{eqn:Tex}} \label{app:HamiltonianforPhiT}
In this section we will derive~\eqref{eqn:Phiex} and~\eqref{eqn:Tex}, which, in the Hamiltonian formalism,  correspond to the matrix elements
\begin{align}\label{eqn:gammaphiTdef}
&\gamma_{[\Phi,T]_J}\Big|_\Phi=f_{\phi\phi t}^2\langle[\Phi,T]_J|\int d^{d+1}x(\phi\phi t^\dagger+\phi^\dagger\phi^\dagger t)|[\Phi^*,T^*]_J\rangle\, , \notag\\
&\gamma_{[\Phi,T]_J}\Big|_S=f_{s\phi\phi^\dagger }f_{stt^\dagger}\langle[\Phi,T]_J| \int d^{d+1}x(\phi\phi^\dagger s+t t^\dagger s)|[\Phi^*,T^*]_J\rangle\, , 
\end{align}
where $f_{\CO_1\CO_2\CO_3}$ are AdS scalar bulk couplings. They are related to CFT OPE coefficients by~\cite{Costa:2014kfa}
\begin{equation}
f_{\alpha\alpha\beta}=c_{\alpha\alpha\beta}\frac{4\sqrt{2}\pi^{\frac{d}{4}}\Gamma(\Delta_\alpha)\Gamma(-\frac{d}{2}+\Delta_\alpha+1)\sqrt{\Gamma(\Delta_\beta)\Gamma(-\frac{d}{2}+\Delta_\beta+1)}}{\Gamma(\frac{\Delta_\beta}{2})^2\Gamma(\Delta_\alpha-\frac{\Delta_\beta}{2})\Gamma(\frac{\Delta_\beta-d}{2}+\Delta_\alpha)}\, .
\end{equation}
The decomposition of double-twist operator $|[\Phi,T]_J\rangle$ into monomial states are given by
\begin{equation}
|[\Phi,T]_J\rangle=\sum_{m=0}^J(-1)^m\sqrt{\frac{\Gamma(1+J)\Gamma(\Delta_T+J)\Gamma(\Delta_\Phi+J)\Gamma(\Delta_T+\Delta_\Phi+J-1)}{m!(J-m)!\Gamma(\Delta_T+m)\Gamma(\Delta_\Phi+J-m)\Gamma(2J-1+\Delta_T+\Delta_\Phi)}}(a_T)^\dagger_{m}(a_\Phi)^\dagger_{J-m}|\text{vac}\rangle\, , 
\end{equation}
where $(a_T)^\dagger$ and $(a_\Phi)^\dagger$ are creating $T$ and $\Phi$ particle respectively.  The monomial matrix elements are derived in detail in~\cite{CompanionPaper1}; here, we simply present the final results for the $\phi$- and $s$-exchange contributions.
\begin{small}
\begin{align}\label{eqn:gammaphiTgen}
&\frac{\gamma_{[\Phi,T]_J}\Big|_{\Phi}}{f_{\phi\phi t}^2}=\frac{\pi^{1-\frac{d}{2}}\Gamma(\frac{\Delta_T}{2})\Gamma(\Delta_\Phi-\frac{\Delta_T}{2})\Gamma(\Delta_\Phi+\frac{\Delta_T-d}{2})^2}{8\sin(\frac{\pi\Delta_T}{2})\Gamma(1-\frac{d}{2}+\Delta_T)\Gamma(1-\frac{d}{2}+\Delta_\Phi)}\notag\\
&\times{}_4\tilde{F}_3\Big[\{1-\frac{\Delta_T}{2},\frac{\Delta_T}{2},\Delta_\Phi-\frac{\Delta_T}{2},\frac{\Delta_T-d}{2}+\Delta_\Phi\},\{1-J-\frac{\Delta_T}{2},1-\frac{d}{2}+\Delta_\Phi,J+\frac{\Delta_T}{2}+\Delta_\Phi\},1\Big]\notag\\
&-\frac{\pi^{\frac{d}{2}}\Gamma(1+J)\Gamma(\frac{\Delta_T}{2})\Gamma(\Delta_T+J)\Gamma(\Delta_\Phi+J)\Gamma(-\frac{d}{2}+J+\Delta_T+\Delta_\Phi)\Gamma(\frac{\Delta_T-d}{2}+\Delta_\Phi)}{8\Gamma(1-\frac{d}{2}+\Delta_T)\Gamma(1-\frac{d}{2}+\Delta_\Phi)}\notag\\
&\times{}_4\tilde{F}_3\Big[\{1+J,\Delta_T+J,\Delta_\Phi+J,-\frac{d}{2}+J+\Delta_T+\Delta_\Phi\},\{1+J+\frac{\Delta_T}{2},1-\frac{d}{2}+J+\frac{\Delta_T}{2}+\Delta_\Phi,2J+\Delta_T+\Delta_\Phi\},1\Big]\, , \notag\\
&\frac{\gamma_{[\Phi,T]_J}\Big|_{S}}{f_{ s\phi\phi^\dagger}f_{s t t^\dagger}}=-(-1)^J\frac{\pi^{1-d/2}\Gamma(\frac{\Delta_S}{2})^2\Gamma(\Delta_\Phi+J)\Gamma(\frac{\Delta_S-d}{2}+\Delta_T)\Gamma(\frac{\Delta_S-d}{2}+\Delta_\Phi)}{16\sin[\frac{\pi(\Delta_S-2\Delta_T)}{2}]\Gamma(-\frac{d}{2}+\Delta_T+1)\Gamma(-\frac{d}{2}+\Delta_\Phi+1)\Gamma(\Delta_T+J)}\notag\\
&\times{}_4\tilde{F}_3\Big[\{\frac{\Delta_S}{2},\frac{\Delta_S}{2},\frac{\Delta_S-2\Delta_T+2}{2},\frac{\Delta_S-d}{2}+\Delta_\Phi\},\{-\frac{d}{2}+\Delta_S+1,\frac{-2J+\Delta_S-2\Delta_T+2}{2},J+\frac{\Delta_S}{2}+\Delta_\Phi\},1\Big]\notag\\
&-(-1)^J\frac{\pi^{-\frac{d}{2}}\Gamma(\Delta_T)^2\Gamma(\Delta_T-\frac{\Delta_S}{2})\Gamma(\Delta_\Phi+J)\Gamma(\frac{\Delta_S-d}{2}+\Delta_T)\Gamma(-\frac{d}{2}+\Delta_T+\Delta_\Phi)}{16\Gamma(-\frac{d}{2}+\Delta_T+1)\Gamma(-\frac{d}{2}+\Delta_\Phi+1)\Gamma(\Delta_T+J)}\notag\\
&\times{}_5\tilde{F}_4\Big[\{1,1,\Delta_T,\Delta_T,-\frac{d}{2}+\Delta_T+\Delta_\Phi\},\{1-J,\frac{-\Delta_S}{2}+\Delta_T+1,\frac{-d+\Delta_S+2}{2}+\Delta_T,J+\Delta_T+\Delta_\Phi\},1\Big]\,  , 
\end{align}
\end{small}
$\!\!$where ${}_p\tilde{F}_q[\{\dots\},\{\dots\},1]$ is the regularized hypergeometric function. Finally, by substituting the explicit OPE data, we obtain the results in~\eqref{eqn:Phiex} and~\eqref{eqn:Tex}.

\twocolumngrid

\bibliographystyle{apsrev4-1-nospace}
\bibliography{refs}

\begin{thebibliography}{28}%
\makeatletter
\providecommand \@ifxundefined [1]{%
 \@ifx{#1\undefined}
}%
\providecommand \@ifnum [1]{%
 \ifnum #1\expandafter \@firstoftwo
 \else \expandafter \@secondoftwo
 \fi
}%
\providecommand \@ifx [1]{%
 \ifx #1\expandafter \@firstoftwo
 \else \expandafter \@secondoftwo
 \fi
}%
\providecommand \natexlab [1]{#1}%
\providecommand \enquote  [1]{``#1''}%
\providecommand \bibnamefont  [1]{#1}%
\providecommand \bibfnamefont [1]{#1}%
\providecommand \citenamefont [1]{#1}%
\providecommand \href@noop [0]{\@secondoftwo}%
\providecommand \href [0]{\begingroup \@sanitize@url \@href}%
\providecommand \@href[1]{\@@startlink{#1}\@@href}%
\providecommand \@@href[1]{\endgroup#1\@@endlink}%
\providecommand \@sanitize@url [0]{\catcode `\\12\catcode `\$12\catcode
  `\&12\catcode `\#12\catcode `\^12\catcode `\_12\catcode `\%12\relax}%
\providecommand \@@startlink[1]{}%
\providecommand \@@endlink[0]{}%
\providecommand \url  [0]{\begingroup\@sanitize@url \@url }%
\providecommand \@url [1]{\endgroup\@href {#1}{\urlprefix }}%
\providecommand \urlprefix  [0]{URL }%
\providecommand \Eprint [0]{\href }%
\providecommand \doibase [0]{http://dx.doi.org/}%
\providecommand \selectlanguage [0]{\@gobble}%
\providecommand \bibinfo  [0]{\@secondoftwo}%
\providecommand \bibfield  [0]{\@secondoftwo}%
\providecommand \translation [1]{[#1]}%
\providecommand \BibitemOpen [0]{}%
\providecommand \bibitemStop [0]{}%
\providecommand \bibitemNoStop [0]{.\EOS\space}%
\providecommand \EOS [0]{\spacefactor3000\relax}%
\providecommand \BibitemShut  [1]{\csname bibitem#1\endcsname}%
\let\auto@bib@innerbib\@empty
\bibitem [{\citenamefont {Fitzpatrick}\ \emph {et~al.}(2013)\citenamefont
  {Fitzpatrick}, \citenamefont {Kaplan}, \citenamefont {Poland},\ and\
  \citenamefont {Simmons-Duffin}}]{Fitzpatrick:2012yx}%
  \BibitemOpen
  \bibfield  {author} {\bibinfo {author} {\bibfnamefont {A.~L.}\ \bibnamefont
  {Fitzpatrick}}, \bibinfo {author} {\bibfnamefont {J.}~\bibnamefont {Kaplan}},
  \bibinfo {author} {\bibfnamefont {D.}~\bibnamefont {Poland}},\ and\ \bibinfo
  {author} {\bibfnamefont {D.}~\bibnamefont {Simmons-Duffin}},\ }\href
  {\doibase 10.1007/JHEP12(2013)004} {\bibfield  {journal} {\bibinfo  {journal}
  {JHEP}\ }\textbf {\bibinfo {volume} {12}},\ \bibinfo {pages} {004} (\bibinfo
  {year} {2013})},\ \Eprint {http://arxiv.org/abs/1212.3616} {arXiv:1212.3616
  [hep-th]} \BibitemShut {NoStop}%
\bibitem [{\citenamefont {Komargodski}\ and\ \citenamefont
  {Zhiboedov}(2013)}]{Komargodski:2012ek}%
  \BibitemOpen
  \bibfield  {author} {\bibinfo {author} {\bibfnamefont {Z.}~\bibnamefont
  {Komargodski}}\ and\ \bibinfo {author} {\bibfnamefont {A.}~\bibnamefont
  {Zhiboedov}},\ }\href {\doibase 10.1007/JHEP11(2013)140} {\bibfield
  {journal} {\bibinfo  {journal} {JHEP}\ }\textbf {\bibinfo {volume} {11}},\
  \bibinfo {pages} {140} (\bibinfo {year} {2013})},\ \Eprint
  {http://arxiv.org/abs/1212.4103} {arXiv:1212.4103 [hep-th]} \BibitemShut
  {NoStop}%
\bibitem [{\citenamefont {Pal}\ \emph {et~al.}(2023)\citenamefont {Pal},
  \citenamefont {Qiao},\ and\ \citenamefont {Rychkov}}]{Pal:2022vqc}%
  \BibitemOpen
  \bibfield  {author} {\bibinfo {author} {\bibfnamefont {S.}~\bibnamefont
  {Pal}}, \bibinfo {author} {\bibfnamefont {J.}~\bibnamefont {Qiao}},\ and\
  \bibinfo {author} {\bibfnamefont {S.}~\bibnamefont {Rychkov}},\ }\href
  {\doibase 10.1007/s00220-023-04767-w} {\bibfield  {journal} {\bibinfo
  {journal} {Commun. Math. Phys.}\ }\textbf {\bibinfo {volume} {402}},\
  \bibinfo {pages} {2169} (\bibinfo {year} {2023})},\ \Eprint
  {http://arxiv.org/abs/2212.04893} {arXiv:2212.04893 [hep-th]} \BibitemShut
  {NoStop}%
\bibitem [{\citenamefont {van Rees}(2024)}]{vanRees:2024xkb}%
  \BibitemOpen
  \bibfield  {author} {\bibinfo {author} {\bibfnamefont {B.~C.}\ \bibnamefont
  {van Rees}},\ }\href@noop {} {\  (\bibinfo {year} {2024})},\ \Eprint
  {http://arxiv.org/abs/2412.06907} {arXiv:2412.06907 [hep-th]} \BibitemShut
  {NoStop}%
\bibitem [{\citenamefont {Fardelli}\ \emph {et~al.}()\citenamefont {Fardelli},
  \citenamefont {Fitzpatrick},\ and\ \citenamefont {Li}}]{CompanionPaper1}%
  \BibitemOpen
  \bibfield  {author} {\bibinfo {author} {\bibfnamefont {G.}~\bibnamefont
  {Fardelli}}, \bibinfo {author} {\bibfnamefont {A.~L.}\ \bibnamefont
  {Fitzpatrick}},\ and\ \bibinfo {author} {\bibfnamefont {W.}~\bibnamefont
  {Li}},\ }\href@noop {} {\ }\Eprint {http://arxiv.org/abs/{Towards Large-Spin
  Effective Theory I: Three-Particle States in AdS $\phi^4$ Theory}} {{Towards
  Large-Spin Effective Theory I: Three-Particle States in AdS $\phi^4$ Theory}}
  \BibitemShut {NoStop}%
\bibitem [{\citenamefont {Kehrein}\ \emph {et~al.}(1993)\citenamefont
  {Kehrein}, \citenamefont {Wegner},\ and\ \citenamefont
  {Pismak}}]{Kehrein:1992fn}%
  \BibitemOpen
  \bibfield  {author} {\bibinfo {author} {\bibfnamefont {S.}~\bibnamefont
  {Kehrein}}, \bibinfo {author} {\bibfnamefont {F.}~\bibnamefont {Wegner}},\
  and\ \bibinfo {author} {\bibfnamefont {Y.}~\bibnamefont {Pismak}},\ }\href
  {\doibase 10.1016/0550-3213(93)90124-8} {\bibfield  {journal} {\bibinfo
  {journal} {Nucl. Phys. B}\ }\textbf {\bibinfo {volume} {402}},\ \bibinfo
  {pages} {669} (\bibinfo {year} {1993})}\BibitemShut {NoStop}%
\bibitem [{\citenamefont {Kehrein}\ and\ \citenamefont
  {Wegner}(1994)}]{Kehrein:1994ff}%
  \BibitemOpen
  \bibfield  {author} {\bibinfo {author} {\bibfnamefont {S.~K.}\ \bibnamefont
  {Kehrein}}\ and\ \bibinfo {author} {\bibfnamefont {F.}~\bibnamefont
  {Wegner}},\ }\href {\doibase 10.1016/0550-3213(94)90406-5} {\bibfield
  {journal} {\bibinfo  {journal} {Nucl. Phys. B}\ }\textbf {\bibinfo {volume}
  {424}},\ \bibinfo {pages} {521} (\bibinfo {year} {1994})},\ \Eprint
  {http://arxiv.org/abs/hep-th/9405123} {arXiv:hep-th/9405123} \BibitemShut
  {NoStop}%
\bibitem [{\citenamefont {Kehrein}(1995)}]{Kehrein:1995ia}%
  \BibitemOpen
  \bibfield  {author} {\bibinfo {author} {\bibfnamefont {S.~K.}\ \bibnamefont
  {Kehrein}},\ }\href {\doibase 10.1016/0550-3213(95)00375-3} {\bibfield
  {journal} {\bibinfo  {journal} {Nucl. Phys. B}\ }\textbf {\bibinfo {volume}
  {453}},\ \bibinfo {pages} {777} (\bibinfo {year} {1995})},\ \Eprint
  {http://arxiv.org/abs/hep-th/9507044} {arXiv:hep-th/9507044} \BibitemShut
  {NoStop}%
\bibitem [{\citenamefont {Derkachov}\ and\ \citenamefont
  {Manashov}(1995{\natexlab{a}})}]{Derkachov:1995zr}%
  \BibitemOpen
  \bibfield  {author} {\bibinfo {author} {\bibfnamefont {S.~E.}\ \bibnamefont
  {Derkachov}}\ and\ \bibinfo {author} {\bibfnamefont {A.~N.}\ \bibnamefont
  {Manashov}},\ }\href {\doibase 10.1016/0550-3213(95)00513-R} {\bibfield
  {journal} {\bibinfo  {journal} {Nucl. Phys. B}\ }\textbf {\bibinfo {volume}
  {455}},\ \bibinfo {pages} {685} (\bibinfo {year} {1995}{\natexlab{a}})},\
  \Eprint {http://arxiv.org/abs/hep-th/9505110} {arXiv:hep-th/9505110}
  \BibitemShut {NoStop}%
\bibitem [{\citenamefont {Derkachov}\ and\ \citenamefont
  {Manashov}(1995{\natexlab{b}})}]{DERKACHOV1995685}%
  \BibitemOpen
  \bibfield  {author} {\bibinfo {author} {\bibfnamefont {S.}~\bibnamefont
  {Derkachov}}\ and\ \bibinfo {author} {\bibfnamefont {A.}~\bibnamefont
  {Manashov}},\ }\href {\doibase https://doi.org/10.1016/0550-3213(95)00513-R}
  {\bibfield  {journal} {\bibinfo  {journal} {Nuclear Physics B}\ }\textbf
  {\bibinfo {volume} {455}},\ \bibinfo {pages} {685} (\bibinfo {year}
  {1995}{\natexlab{b}})}\BibitemShut {NoStop}%
\bibitem [{\citenamefont {Liendo}(2017)}]{Liendo:2017wsn}%
  \BibitemOpen
  \bibfield  {author} {\bibinfo {author} {\bibfnamefont {P.}~\bibnamefont
  {Liendo}},\ }\href {\doibase 10.1016/j.nuclphysb.2017.04.020} {\bibfield
  {journal} {\bibinfo  {journal} {Nucl. Phys. B}\ }\textbf {\bibinfo {volume}
  {920}},\ \bibinfo {pages} {368} (\bibinfo {year} {2017})},\ \Eprint
  {http://arxiv.org/abs/1701.04830} {arXiv:1701.04830 [hep-th]} \BibitemShut
  {NoStop}%
\bibitem [{\citenamefont {Wilson}\ and\ \citenamefont
  {Kogut}(1974)}]{Wilson:1973jj}%
  \BibitemOpen
  \bibfield  {author} {\bibinfo {author} {\bibfnamefont {K.~G.}\ \bibnamefont
  {Wilson}}\ and\ \bibinfo {author} {\bibfnamefont {J.~B.}\ \bibnamefont
  {Kogut}},\ }\href {\doibase 10.1016/0370-1573(74)90023-4} {\bibfield
  {journal} {\bibinfo  {journal} {Phys. Rept.}\ }\textbf {\bibinfo {volume}
  {12}},\ \bibinfo {pages} {75} (\bibinfo {year} {1974})}\BibitemShut {NoStop}%
\bibitem [{\citenamefont {Caron-Huot}(2017)}]{Caron-Huot:2017vep}%
  \BibitemOpen
  \bibfield  {author} {\bibinfo {author} {\bibfnamefont {S.}~\bibnamefont
  {Caron-Huot}},\ }\href {\doibase 10.1007/JHEP09(2017)078} {\bibfield
  {journal} {\bibinfo  {journal} {JHEP}\ }\textbf {\bibinfo {volume} {09}},\
  \bibinfo {pages} {078} (\bibinfo {year} {2017})},\ \Eprint
  {http://arxiv.org/abs/1703.00278} {arXiv:1703.00278 [hep-th]} \BibitemShut
  {NoStop}%
\bibitem [{\citenamefont {Simmons-Duffin}(2017)}]{Simmons-Duffin:2016wlq}%
  \BibitemOpen
  \bibfield  {author} {\bibinfo {author} {\bibfnamefont {D.}~\bibnamefont
  {Simmons-Duffin}},\ }\href {\doibase 10.1007/JHEP03(2017)086} {\bibfield
  {journal} {\bibinfo  {journal} {JHEP}\ }\textbf {\bibinfo {volume} {03}},\
  \bibinfo {pages} {086} (\bibinfo {year} {2017})},\ \Eprint
  {http://arxiv.org/abs/1612.08471} {arXiv:1612.08471 [hep-th]} \BibitemShut
  {NoStop}%
\bibitem [{\citenamefont {Penedones}(2011)}]{Penedones:2010ue}%
  \BibitemOpen
  \bibfield  {author} {\bibinfo {author} {\bibfnamefont {J.}~\bibnamefont
  {Penedones}},\ }\href {\doibase 10.1007/JHEP03(2011)025} {\bibfield
  {journal} {\bibinfo  {journal} {JHEP}\ }\textbf {\bibinfo {volume} {03}},\
  \bibinfo {pages} {025} (\bibinfo {year} {2011})},\ \Eprint
  {http://arxiv.org/abs/1011.1485} {arXiv:1011.1485 [hep-th]} \BibitemShut
  {NoStop}%
\bibitem [{\citenamefont {Anand}\ \emph {et~al.}(2020)\citenamefont {Anand},
  \citenamefont {Fitzpatrick}, \citenamefont {Katz}, \citenamefont {Khandker},
  \citenamefont {Walters},\ and\ \citenamefont {Xin}}]{Anand:2020gnn}%
  \BibitemOpen
  \bibfield  {author} {\bibinfo {author} {\bibfnamefont {N.}~\bibnamefont
  {Anand}}, \bibinfo {author} {\bibfnamefont {A.~L.}\ \bibnamefont
  {Fitzpatrick}}, \bibinfo {author} {\bibfnamefont {E.}~\bibnamefont {Katz}},
  \bibinfo {author} {\bibfnamefont {Z.~U.}\ \bibnamefont {Khandker}}, \bibinfo
  {author} {\bibfnamefont {M.~T.}\ \bibnamefont {Walters}},\ and\ \bibinfo
  {author} {\bibfnamefont {Y.}~\bibnamefont {Xin}},\ }\href@noop {} {\
  (\bibinfo {year} {2020})},\ \Eprint {http://arxiv.org/abs/2005.13544}
  {arXiv:2005.13544 [hep-th]} \BibitemShut {NoStop}%
\bibitem [{\citenamefont {Bertucci}\ \emph {et~al.}(2022)\citenamefont
  {Bertucci}, \citenamefont {Henriksson},\ and\ \citenamefont
  {McPeak}}]{Bertucci:2022ptt}%
  \BibitemOpen
  \bibfield  {author} {\bibinfo {author} {\bibfnamefont {F.}~\bibnamefont
  {Bertucci}}, \bibinfo {author} {\bibfnamefont {J.}~\bibnamefont
  {Henriksson}},\ and\ \bibinfo {author} {\bibfnamefont {B.}~\bibnamefont
  {McPeak}},\ }\href {\doibase 10.1007/JHEP10(2022)104} {\bibfield  {journal}
  {\bibinfo  {journal} {JHEP}\ }\textbf {\bibinfo {volume} {10}},\ \bibinfo
  {pages} {104} (\bibinfo {year} {2022})},\ \Eprint
  {http://arxiv.org/abs/2205.09132} {arXiv:2205.09132 [hep-th]} \BibitemShut
  {NoStop}%
\bibitem [{\citenamefont {Alday}\ \emph {et~al.}(2018)\citenamefont {Alday},
  \citenamefont {Henriksson},\ and\ \citenamefont {van Loon}}]{Alday:2017zzv}%
  \BibitemOpen
  \bibfield  {author} {\bibinfo {author} {\bibfnamefont {L.~F.}\ \bibnamefont
  {Alday}}, \bibinfo {author} {\bibfnamefont {J.}~\bibnamefont {Henriksson}},\
  and\ \bibinfo {author} {\bibfnamefont {M.}~\bibnamefont {van Loon}},\ }\href
  {\doibase 10.1007/JHEP07(2018)131} {\bibfield  {journal} {\bibinfo  {journal}
  {JHEP}\ }\textbf {\bibinfo {volume} {07}},\ \bibinfo {pages} {131} (\bibinfo
  {year} {2018})},\ \Eprint {http://arxiv.org/abs/1712.02314} {arXiv:1712.02314
  [hep-th]} \BibitemShut {NoStop}%
\bibitem [{\citenamefont {Henriksson}\ and\ \citenamefont
  {Van~Loon}(2019)}]{Henriksson:2018myn}%
  \BibitemOpen
  \bibfield  {author} {\bibinfo {author} {\bibfnamefont {J.}~\bibnamefont
  {Henriksson}}\ and\ \bibinfo {author} {\bibfnamefont {M.}~\bibnamefont
  {Van~Loon}},\ }\href {\doibase 10.1088/1751-8121/aaf1e2} {\bibfield
  {journal} {\bibinfo  {journal} {J. Phys. A}\ }\textbf {\bibinfo {volume}
  {52}},\ \bibinfo {pages} {025401} (\bibinfo {year} {2019})},\ \Eprint
  {http://arxiv.org/abs/1801.03512} {arXiv:1801.03512 [hep-th]} \BibitemShut
  {NoStop}%
\bibitem [{\citenamefont {Henriksson}(2023)}]{Henriksson:2022rnm}%
  \BibitemOpen
  \bibfield  {author} {\bibinfo {author} {\bibfnamefont {J.}~\bibnamefont
  {Henriksson}},\ }\href {\doibase 10.1016/j.physrep.2022.12.002} {\bibfield
  {journal} {\bibinfo  {journal} {Phys. Rept.}\ }\textbf {\bibinfo {volume}
  {1002}},\ \bibinfo {pages} {1} (\bibinfo {year} {2023})},\ \Eprint
  {http://arxiv.org/abs/2201.09520} {arXiv:2201.09520 [hep-th]} \BibitemShut
  {NoStop}%
\bibitem [{\citenamefont {Gopakumar}\ \emph
  {et~al.}(2017{\natexlab{a}})\citenamefont {Gopakumar}, \citenamefont
  {Kaviraj}, \citenamefont {Sen},\ and\ \citenamefont
  {Sinha}}]{Gopakumar:2016wkt}%
  \BibitemOpen
  \bibfield  {author} {\bibinfo {author} {\bibfnamefont {R.}~\bibnamefont
  {Gopakumar}}, \bibinfo {author} {\bibfnamefont {A.}~\bibnamefont {Kaviraj}},
  \bibinfo {author} {\bibfnamefont {K.}~\bibnamefont {Sen}},\ and\ \bibinfo
  {author} {\bibfnamefont {A.}~\bibnamefont {Sinha}},\ }\href {\doibase
  10.1103/PhysRevLett.118.081601} {\bibfield  {journal} {\bibinfo  {journal}
  {Phys. Rev. Lett.}\ }\textbf {\bibinfo {volume} {118}},\ \bibinfo {pages}
  {081601} (\bibinfo {year} {2017}{\natexlab{a}})},\ \Eprint
  {http://arxiv.org/abs/1609.00572} {arXiv:1609.00572 [hep-th]} \BibitemShut
  {NoStop}%
\bibitem [{\citenamefont {Gopakumar}\ \emph
  {et~al.}(2017{\natexlab{b}})\citenamefont {Gopakumar}, \citenamefont
  {Kaviraj}, \citenamefont {Sen},\ and\ \citenamefont
  {Sinha}}]{Gopakumar:2016cpb}%
  \BibitemOpen
  \bibfield  {author} {\bibinfo {author} {\bibfnamefont {R.}~\bibnamefont
  {Gopakumar}}, \bibinfo {author} {\bibfnamefont {A.}~\bibnamefont {Kaviraj}},
  \bibinfo {author} {\bibfnamefont {K.}~\bibnamefont {Sen}},\ and\ \bibinfo
  {author} {\bibfnamefont {A.}~\bibnamefont {Sinha}},\ }\href {\doibase
  10.1007/JHEP05(2017)027} {\bibfield  {journal} {\bibinfo  {journal} {JHEP}\
  }\textbf {\bibinfo {volume} {05}},\ \bibinfo {pages} {027} (\bibinfo {year}
  {2017}{\natexlab{b}})},\ \Eprint {http://arxiv.org/abs/1611.08407}
  {arXiv:1611.08407 [hep-th]} \BibitemShut {NoStop}%
\bibitem [{\citenamefont {Dey}\ \emph {et~al.}(2017)\citenamefont {Dey},
  \citenamefont {Kaviraj},\ and\ \citenamefont {Sinha}}]{Dey:2016mcs}%
  \BibitemOpen
  \bibfield  {author} {\bibinfo {author} {\bibfnamefont {P.}~\bibnamefont
  {Dey}}, \bibinfo {author} {\bibfnamefont {A.}~\bibnamefont {Kaviraj}},\ and\
  \bibinfo {author} {\bibfnamefont {A.}~\bibnamefont {Sinha}},\ }\href
  {\doibase 10.1007/JHEP07(2017)019} {\bibfield  {journal} {\bibinfo  {journal}
  {JHEP}\ }\textbf {\bibinfo {volume} {07}},\ \bibinfo {pages} {019} (\bibinfo
  {year} {2017})},\ \Eprint {http://arxiv.org/abs/1612.05032} {arXiv:1612.05032
  [hep-th]} \BibitemShut {NoStop}%
\bibitem [{\citenamefont {Dey}\ and\ \citenamefont
  {Kaviraj}(2018)}]{Dey:2017oim}%
  \BibitemOpen
  \bibfield  {author} {\bibinfo {author} {\bibfnamefont {P.}~\bibnamefont
  {Dey}}\ and\ \bibinfo {author} {\bibfnamefont {A.}~\bibnamefont {Kaviraj}},\
  }\href {\doibase 10.1007/JHEP02(2018)153} {\bibfield  {journal} {\bibinfo
  {journal} {JHEP}\ }\textbf {\bibinfo {volume} {02}},\ \bibinfo {pages} {153}
  (\bibinfo {year} {2018})},\ \Eprint {http://arxiv.org/abs/1711.01173}
  {arXiv:1711.01173 [hep-th]} \BibitemShut {NoStop}%
\bibitem [{\citenamefont {Henriksson}\ \emph {et~al.}(2024)\citenamefont
  {Henriksson}, \citenamefont {Kravchuk},\ and\ \citenamefont
  {Oertel}}]{Henriksson:2023cnh}%
  \BibitemOpen
  \bibfield  {author} {\bibinfo {author} {\bibfnamefont {J.}~\bibnamefont
  {Henriksson}}, \bibinfo {author} {\bibfnamefont {P.}~\bibnamefont
  {Kravchuk}},\ and\ \bibinfo {author} {\bibfnamefont {B.}~\bibnamefont
  {Oertel}},\ }\href {\doibase 10.1007/JHEP07(2024)248} {\bibfield  {journal}
  {\bibinfo  {journal} {JHEP}\ }\textbf {\bibinfo {volume} {07}},\ \bibinfo
  {pages} {248} (\bibinfo {year} {2024})},\ \Eprint
  {http://arxiv.org/abs/2312.09283} {arXiv:2312.09283 [hep-th]} \BibitemShut
  {NoStop}%
\bibitem [{\citenamefont {Fardelli}\ \emph {et~al.}(2024)\citenamefont
  {Fardelli}, \citenamefont {Fitzpatrick},\ and\ \citenamefont
  {Li}}]{Fardelli:2024heb}%
  \BibitemOpen
  \bibfield  {author} {\bibinfo {author} {\bibfnamefont {G.}~\bibnamefont
  {Fardelli}}, \bibinfo {author} {\bibfnamefont {A.~L.}\ \bibnamefont
  {Fitzpatrick}},\ and\ \bibinfo {author} {\bibfnamefont {W.}~\bibnamefont
  {Li}},\ }\href {\doibase 10.1007/JHEP08(2024)202} {\bibfield  {journal}
  {\bibinfo  {journal} {JHEP}\ }\textbf {\bibinfo {volume} {08}},\ \bibinfo
  {pages} {202} (\bibinfo {year} {2024})},\ \Eprint
  {http://arxiv.org/abs/2403.07079} {arXiv:2403.07079 [hep-th]} \BibitemShut
  {NoStop}%
\bibitem [{\citenamefont {Fitzpatrick}\ and\ \citenamefont
  {Kaplan}(2012)}]{Fitzpatrick:2011dm}%
  \BibitemOpen
  \bibfield  {author} {\bibinfo {author} {\bibfnamefont {A.~L.}\ \bibnamefont
  {Fitzpatrick}}\ and\ \bibinfo {author} {\bibfnamefont {J.}~\bibnamefont
  {Kaplan}},\ }\href {\doibase 10.1007/JHEP10(2012)032} {\bibfield  {journal}
  {\bibinfo  {journal} {JHEP}\ }\textbf {\bibinfo {volume} {10}},\ \bibinfo
  {pages} {032} (\bibinfo {year} {2012})},\ \Eprint
  {http://arxiv.org/abs/1112.4845} {arXiv:1112.4845 [hep-th]} \BibitemShut
  {NoStop}%
\bibitem [{\citenamefont {Costa}\ \emph {et~al.}(2014)\citenamefont {Costa},
  \citenamefont {Gon\c{c}alves},\ and\ \citenamefont
  {Penedones}}]{Costa:2014kfa}%
  \BibitemOpen
  \bibfield  {author} {\bibinfo {author} {\bibfnamefont {M.~S.}\ \bibnamefont
  {Costa}}, \bibinfo {author} {\bibfnamefont {V.}~\bibnamefont
  {Gon\c{c}alves}},\ and\ \bibinfo {author} {\bibfnamefont {J.~a.}\
  \bibnamefont {Penedones}},\ }\href {\doibase 10.1007/JHEP09(2014)064}
  {\bibfield  {journal} {\bibinfo  {journal} {JHEP}\ }\textbf {\bibinfo
  {volume} {09}},\ \bibinfo {pages} {064} (\bibinfo {year} {2014})},\ \Eprint
  {http://arxiv.org/abs/1404.5625} {arXiv:1404.5625 [hep-th]} \BibitemShut
  {NoStop}%
\end{thebibliography}%

\end{document}